\documentclass[
reprint,
superscriptaddress,
showpacs,
amsmath,amssymb,
aps,
prb,
floatfix
]{revtex4-1}
\usepackage{float}
\usepackage{graphicx}
\usepackage{dcolumn}
\usepackage{bm}

\begin{document}
	\date{\today}
	\title{Devil's staircase structures in $\varphi_{0}$ junction}
	\author{M. Nashaat}
	\email{majed@sci.cu.edu.eg}
	\affiliation{Department of Physics, Cairo University, Cairo, 12613, Egypt}
	\affiliation{BLTP, JINR, Dubna, Moscow Region, 141980, Russia}
	\author{Yu. M. Shukrinov}
	\affiliation{BLTP, JINR, Dubna, Moscow Region, 141980, Russia}
	\affiliation{Dubna State University, Dubna, Russian Federation}
	\author{A.Y. Ellithi }
	\affiliation{Department of Physics, Cairo University, Cairo, 12613, Egypt}
	\author{Th. M. El Sherbini}
	\affiliation{Department of Physics, Cairo University, Cairo, 12613, Egypt}
           
	\vspace{10pt}

\begin{abstract}
	
The superconductor-ferromagnet-superconductor $\varphi_{0}$ junction provides a direct coupling between Josephson phase and magnetic moment of ferromagnetic barrier. We demonstrate an appearance of additional fractional subharmonic steps in the IV-characteristics of $\varphi_{0}$ junction under external electromagnetic radiation due to spin-orbit coupling. An origin of subharmonic steps is related to the locking of magnetic moment precession to the Josephson oscillations. We prove that the positions of those steps follow a continued fraction algorithm.

\end{abstract}

\maketitle

\section{Introduction}
Josephson junction (JJ) with ferromagnetic barrier opens an additional gate for investigation of the interplay between ferromagnetism and  superconductivity. Its  current-phase relation is very sensitive to the characteristic of the ferromagnet \cite{	Braude2007,buzdin2008direct,Linder2015,Silaev2017,Bobkova2017,Rabinovich2018}.  Particularly, the superconductor-ferromagnet-superconductor (SFS) Josephson junction oscillates between
	$0$- and $\pi$ states with an increase in ferromagnetic layer thickness  \cite{	Ryazanov2001,Oboznov2006,Robinson2006,Iver2008}. SFS JJ holds  a great possibility for the  realization  of  the  qubit  for  quantum  computation \cite{Yamashita2005,Massarotti2015,Yamashita2017} and  can be used as electronic and spintronic components, such as phase shifters and superconducting spin valves \cite{Golod2010,	Leksin2011,Gingrich2016,Soloviev,Zhu2017,Niedzielski2018,Golod2019}.

In case of non-centrosymmetric magnetic  barrier with broken inversion symmetry like $MnSi$ or $FeGe$, the spin-orbit coupling leads to a general current-phase relation $J = J_{c} \sin(\varphi-\varphi_{0})$,
where $\varphi_{0}$ is proportional to the magnetic moment perpendicular to the gradient of the asymmetric spin-orbit potential \cite{buzdin2008direct,buzdin2009}. In this case the spin-orbit coupling in $\varphi_{0}$-junction provides a direct coupling between the suppercurrent and the magnetic moment. This coupling leads to the supercurrent induced magnetization dynamics \cite{buzdin2009,Linder2014,Chudnovsky2016}. Recently, in Ref.\cite{Assouline2019} the authors report the observation of anomalous phase shift $\varphi_{0}$ in $Bi_{2}Se_{3}$ Josephson junctions (JJ) and provide a direct measurement of the spin-orbit coupling strength. Also, in $\varphi_{0}$ junction a full magnetization reversal is demonstrated by using electric current pulse \cite{shukrinov2017magnetization}.

An increase of the Shapiro steps amplitude due to spin-orbit coupling near the ferromagnetic resonance,  the appearance of the half-integer Shapiro steps, and precession of the magnetization vector with radiation frequency \cite{buzdin2009} are predicted, if the $\varphi_0$ Josephson junction is exposed to microwave radiation. An external electromagnetic field can control
qualitative features of the magnetic moment dynamics in a current interval which corresponds to the Shapiro step. The
radiation can also produce topological transformations of precession trajectories~\cite{cond-mat18}, which depend on the amplitude of the applied electromagnetic radiation. Such results might be used for developing novel experimental resonance methods for determination of the spin-orbit interaction in the non-centrosymmetric  materials.

Generally, the IV-characteristic of a conventional JJ in the underdamped case demonstrates harmonic and subharmonic steps in the presence of external electromagnetic radiation\cite{shukrinov2013devil,azbel}, while it shows harmonic steps only in the overdamped junctions \cite{RENNE,WALDRAM}. The steps for the underdamped JJ can form the so-called devil's staircase (DS) structure as a consequence of the interplay between Josephson  and applied frequencies \cite{ben-jacob,shukrinov2013devil,shukrinov2014structured,sokolovic2017devil}.  The DS structure is a universal phenomenon and  appears in a wide variety of different systems, including infinite spin chains with long-range interactions\cite{nebendahl13}, frustrated quasi-two-dimensional spin-dimer systems in magnetic fields\cite{takigawa13}, and even in the fractional quantum Hall effect\cite{hriscu13}. In Ref.\cite{chen2017} a series of fractional integer size steps is observed experimentally in the Kondo lattice CeSbSe.

Coupling between the superconducting current and magnetic moment in the SFS Josephson junction is one of the most interesting topics nowadays \cite{Linder2015}. The connection between the staircase structure and current-phase relation provides an intriguing point in this field of science. Especially, the manifestation of the staircase structure in the IV-characteristics of the junction with the corresponding information on current-phase relation\cite{Golubov,Sellier2004,Maiti2015,Pic2017,Shukrinov2018}, thus, serves as a novel method for its determination. The appearance of the DS structure and its connection with the current-phase relation in experimental situations still in need for detailed investigations. 

In Ref.\cite{Shukrinov2018} we have demonstrated an appearance of subharmonic steps in the IV-characteristic for  overdamped SFS junction with spin wave excitations (magnons). It was found that the width of the current steps at $V=2\Omega$ (where $\Omega$ is the frequency of the applied magnetic field) are larger than the width of the odd and fractional steps. Therefore, magnetization dynamics  can be manifested in the IV-characteristic since this behavior is different from that of the conventional Shapiro step\cite{Shukrinov2018,hikino2011ferromagnetic}. The origin of the even steps is related to the interaction of Cooper pairs with even number of magnons \cite{hikino2011ferromagnetic}.  In Ref.\cite{Mori2014} the authors show that the breathing of the domain wall leads to the appearance of staircase structure in the IV-characteristic of SFS junction, but they did not investigate its relation with DS structure.  The IV-characteristics of superconductor-quantum spin Hall insulator-superconductor system \cite{Hang2018}  in the presence of microwave radiation exhibit a structure where odd steps are completely suppressed, implying a fractional Josephson effect.

Analytic treatment of the coupled JJ-nanomagnet system driven by a time-dependent magnetic field both without and with an external ac field is studied in Ref.\cite{Roopayan2017}. The authors have shown the existence of Shapiro-type steps in the IV-characteristics of the JJ subjected to a voltage bias for a constant or periodically varying magnetic field and explore the effect of rotation of the magnetic field and the presence of an external ac drive on these steps\cite{Roopayan2017}. Here we clarify the possibility for the appearance of  fractional steps  in the overdamped $\varphi_{0}$ junction  using a modified RSJ model.

In this paper,  we focus on the overdamped case ($\beta_{c} = 0$, $\beta_{c}$ is the McCumber parameter)  to reflect clearly the appearance of the subharmonic steps due to the presence of spin-orbit coupling in $\varphi_{0}$ junction. It is found that due to the coupling between Josephson phase and magnetic moment through the spin-orbit coupling, the additional fractional subharmonic steps appear in the IV-characteristic. The appearance and the positions of those steps depend directly on the spin-orbit coupling and the ratio of Josephson and magnetic energies. An analytical consideration of the linearized model justifies the appearance of the fractional steps in IV-characteristics, in agreement with our numerical results.

The plan of the rest part of the paper is as follows. In Sec.I, we describe the self-consistent modified RSJ and LLG dynamical equations for $\varphi_{0}$ junction. This is followed by a discussion of the IV-characteristics of such systems in Sec II. We demonstrate the effect of spin-orbit coupling in the IV-characteristic. Then, we confirm that the appearance of the subharmonic steps under external radiation are due to the presence of spin-orbit coupling.  In addition to this we justify analytically the conditions of the frequency locking and discuss the experimental realization of the found effects. Finally, we conclude in Sec. III. Some details of our calculations are specified in the supplemental material.

\section{Model and Methods}

\begin{figure}[!ht]
	\centering
	\includegraphics[width=0.6\linewidth]{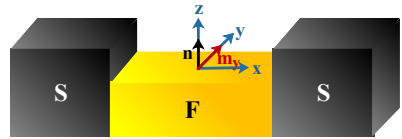}
	\caption{(Color online) Geometry of the $\varphi_{0}-$ junction. $S$ - superconductor, $F$ - ferromagnet, $\textbf{n}$ - unit vector of the gradient of the spin-orbit potential. }
	\label{fig1}
\end{figure}

The geometry of the considered $\varphi_{0}$ junction is shown in figure~\ref{fig1}.  The ferromagnet easy-axis and the gradient of the spin-orbit potential ($n$)  are directed along the z-axis. In this case, $\varphi_{0}= r m_{y}$ and $r$ characterizes the relative strength of the spin-orbit interaction\cite{buzdin2009}.
The electric current through JJs in the dimensionless form is determined by RSJ equation \cite{stewart1968current} (we consider overdamped case with $\beta_{c}=0$  )
\begin{eqnarray}
I + A \sin \Omega t &=&\sin (\varphi -r m_{y}) + \frac{{\rm d}\varphi}{{\rm d}t},
\label{eq1}
\end{eqnarray}
where $I$ is normalized to the critical current $I_{c}$, $t$ is normalized to $\omega^{-1}_{c}$, $\omega_{c}$=$2eRI_{c}/\hbar$, $m_{y}=M_{y}/M_{s}$ satisfies the constraint $\sum_{i=x,y,z} m_{i}^{2}(t)=1$, $M_{s}=\|\textbf{M}\|$,  and the voltage $V$ is normalized to $I_{c} R$. The external
electromagnetic radiation is characterized by the amplitude $I_{ac}$ and frequency $\omega$, $A=I_{ac}/I_{c}$.  The dynamics of the magnetic moment $m_{y}$ is determined by the LLG \cite{ounadjela2003spin}:
{\small
	\begin{eqnarray}
	\frac{d\textbf{m}}{dt} = - \frac{\Omega_{F}}{(1+\alpha^{2})} \bigg(\textbf{m} \times \textbf{}\textbf{h}_{eff} + \alpha \left[ \textbf{m} \times (\textbf{m} \times \textbf{h}_{eff})\right] \bigg), 
	\label{LLG}
	\end{eqnarray}}
where $\Omega_{F}=\omega_{F}/\omega_{c}$, $\omega_{F}$ is the ferromagnetic resonance frequency, $\alpha$ is the Gilbert damping ( here we consider $\alpha=0.1$) and  $\textbf{h}_{eff}$ is the total effective field of the system which is given by $-\boldsymbol{\nabla}_{M} E/v$, $v$ is the volume of ferromagnet. The total energy of the system for the current biased junction is  $E=E_{s} + E_{M}$ with
\begin{eqnarray}
E_{s\;} &=& -\frac{\Phi_0}{2\pi I_{c}} \varphi I + E_{J} \left[1-\cos\left( \varphi -r\frac{M_{y}}{M_{0}} \right)\right],  \nonumber \\
E_{M} &=& - \frac{K v}{2} \bigg(\frac{M_{z}}{M_{s}}\bigg)^{2},
\end{eqnarray}
where $E_{J}= \Phi_{0} I_{c}/2\pi$ is the Josephson energy, $E_{M}$ is the anisotropy energy and $K_{an}=\omega_{F} M_{s}/\gamma$ is the anisotropy constant \cite{buzdin2009,shukrinov2017magnetization}. So, the effective field in normalized form is given by

\begin{equation}
\textbf{h}_{e} =  [ G r \sin(\varphi-r m_{y}) \hat{\textbf{e}_{y}} +m_{z} \hat{\textbf{e}_{z}}],
\end{equation}
with $G=E_{J}/K_{an}v$. The magnetic moment and the phase dynamics of the considered S/F/S Josephson junction is
determined by Eqs.~(\ref{eq1}) and (\ref{LLG}).

To compute the IV-characteristics, we study the temporal dependence of $V(t)=\hbar \dot{\varphi}(t)/(2e)$ obtained by numerical solution of equation (\ref{eq1}). The dc bias current $I$ is  normalized to $I_{c}$, and the voltage is normalized to $\hbar\omega_{c}/(2e)$. We assume a constant bias current and calculate the average voltage. We employ a  fourth-order Runge-Kutta integration scheme. As a result, we find the temporal dependence of the voltage in the junction at a fixed value of bias current $I$. Then, the current value is increased or decreased by a small amount $\delta I$ (the bias current step), to calculate the voltage at the next point of the  IV-characteristics. We use the final voltage achieved at the previous point of the IV-characteristics as the initial condition for the next current point. The average of the voltage $V_{av}$ is given by $V_{av} =\frac{1}{T_{f}-T_{i}}\int^{T_{f}}_{T_{i}}V(t) {\rm d}t$, where $T_{i}$ and $T_{f}$ determine the interval for the temporal averaging. The initial conditions for the magnetization components are assumed to be $m_{x}=0$, $m_{y}=0$ and $m_{z}=1$, while for the voltage and phase we take zeros.

\begin{figure}[!ht]
 	\begin{minipage}{0.5\textwidth}	
 		\centering
 		\includegraphics[width=0.8\linewidth, angle =0]{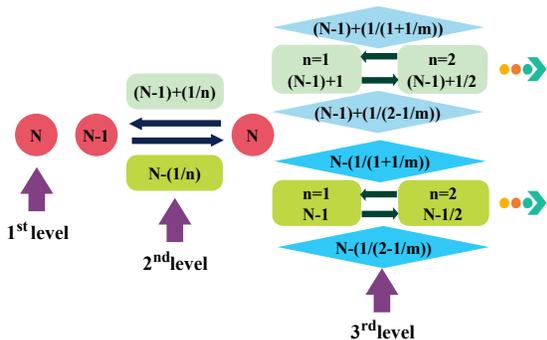}
 		\caption{ Schematic demonstration of the appearance of continued fractions in IV-characteristic of SFS junction under external magnetic field. N is the Shapiro step number, n and m are positive integers \cite{shukrinov2013devil}.}
 		\label{fig:algorithm}
 	\end{minipage}
 \end{figure}

To analyze the positions of subharmonic steps in the IV-characteristics and reflect the DS structure, we use the continued fractions. They consist of number of levels. The first-level is denoted by N and the second-level gives two groups of subharmonics (N-1+1/n) and (N-(1/n)). The algorithm of continued fractions is presented in figure  \ref{fig:algorithm} where the red circles represent the Shapiro step number (first level). The green rectangles represent the second level subharmonic steps and the arrows represent the approaching direction, for example, the group $(N-1)+(1/n)$ approaches $(N-1)^{th}$ Shapiro step and $N-(1/n)$ approaches the $N^{th}$ Shapiro step. The third level subharmonics is represented by the blue diamond. This level is determined by fixing $n$, $n+1$ and changing $m$, for example, the  second level group $(N-1)+(1/n)$ gives rise to two third level groups between $n=1$ and $n=2$, also two other third level group are a rise from the second group $N-(1/n)$ with $n=1$ and $n=2$, etc \cite{shukrinov2013devil,shukrinov2014structured,Shukrinov2018}. The positions of the current steps follow continued fraction formula \cite{shukrinov2013devil},
 \begin{equation}
 V=\left( N \pm \frac{1}{n\pm \frac{1}{m\pm\frac{1}{p\pm...}}}\right) \Omega
 \label{eq:conti}
 \end{equation}
 where N,n,m,p,... are positive integers. Terms with only N form harmonics, while other terms describe subharmonics or fractional steps.

In what follows, we first obtain an approximate analytical solution  of equation (\ref{eq1}) in Sec.B.1,  which  demonstrates  the existence  of  subharmonic  Shapiro  steps  for  $\varphi_{0}$ junction.  Then,  in Sec.B.2, we  carry out a detailed numerical study of equation (\ref{eq1}) where the analytical results  are  verified  and  the  devil’s  staircase  structure  of  the Shapiro steps is studied.

\section{Results and discussion}

\subsection{ Perturbative analytical solution }

First, we find the expression for $m_{y}$. If the deviation of the magnetic moment from the equilibrium point due to Josephson energy is small (i.e., $G<1$), we can linearize the LLG equation. In this case \cite{supplement}, the magnetic moment $m_{y}(t)$ reads
\begin{eqnarray}
m_{y}(t)&\approx&\frac{\tilde{\gamma}_{2}}{D}\sin \varphi(t)-\frac{r\tilde{\gamma}^{2}_{2}}{2D^{2}} \sin 2\varphi(t),
\label{eqa_my}
\end{eqnarray}
where $\tilde{\gamma}_{2}=Gr\gamma_{2}$, $\gamma_{2}=\left( 1-(1-\alpha^{2}) \frac{\Omega_{p}^{2}}{\Omega_{F}^{2}}\right)$, and $D=\left( 1-(1+\alpha^{2})\frac{\Omega_{p}^{2}}{\Omega_{F}^{2}}\right) ^{2} + 4 \alpha^{2}\frac{\Omega_{p}^{2}}{\Omega_{F}^{2}}$.

Then we demonstrate the results of the   perturbative   analysis   of equation (\ref{eq1}) for overdamped $\varphi_{0}$ junctions in high-frequency limit \cite{Kornev}. We analyze this equation for $\Omega$ and $A>>1$  in which we can do the following expansions \cite{Shukrinov2011,Likharev}
\begin{eqnarray}
\varphi(t) = \sum_{n}^{} \epsilon^{n} \varphi_{n}(t), \ \ \  I = \sum_{n=0}^{\infty} \epsilon^{n} I_{n}.
\label{series}
\end{eqnarray}
where $I_{0}$ is the biased current, $\epsilon<<1$ and $I_{n}$ for $n > 0$ are determined self-consistently from the condition of the absence of additional dc voltage: $\lim_{T\longrightarrow\infty}$ $\int_{0}^{T}$ $\dot{\varphi}_{n}dt=0$\cite{Likharev}.  Using equation (\ref{eqa_my}), the RSJ up to terms $\sim O( r m_{y})$ has the form
\begin{eqnarray}
\dot{\varphi}(t)&\approx&I + A \sin \Omega t - \sin \varphi(t)\nonumber \\ &+&\frac{r \tilde{\gamma}_{2}}{2D} \bigg( \sin 2\varphi(t) - \frac{r \tilde{\gamma}_{2}}{2D} \bigg[\sin 3\varphi(t)-\sin\varphi(t)\bigg]\bigg),
\label{eqa1}
\end{eqnarray}
Using (\ref{series}), the equations for $\dot{\varphi}_{n}$ can be obtained by equating terms in the same order of $\epsilon$. Then, the expression of $\dot{\varphi}_{n}$ is given by
\begin{eqnarray}
\dot{\varphi}_{n}(t) =I_{n} +f_{n} (t).
\end{eqnarray}
For  $n=0$ and $n=1$, we have
\begin{eqnarray}
f_{0}(t) &=& A \sin \Omega t, \nonumber \\
f_{1}(t) &=& \bigg[\bigg (\frac{r \tilde{\gamma}_{2}}{2D}\bigg)^{2}-1\bigg]\sin\varphi_{0}(t) +\frac{r \tilde{\gamma}_{2}}{2D} \sin 2\varphi_{0}(t) \nonumber \\ &-& \bigg (\frac{r \tilde{\gamma}_{2}}{2D}\bigg)^{2} \sin 3\varphi_{0}(t),
\end{eqnarray}
The $0^{th}$ order with $n = 0$ represents the autonomous IV-characteristic of the junction. In this case we have
\begin{eqnarray}
\dot{\varphi}_{0}(t) &=&I_{0} + A \sin \Omega t,
\label{phidot_0}
\end{eqnarray}
while the supercurrent is given by
\begin{eqnarray}
I^{(0)}_{s} &=& \bigg[1-\bigg (\frac{r \tilde{\gamma}_{2}}{2D}\bigg)^{2}\bigg]\sin\varphi_{0}(t) -\frac{r \tilde{\gamma}_{2}}{2D} \sin 2\varphi_{0}(t) \nonumber \\ &+& \bigg (\frac{r \tilde{\gamma}_{2}}{2D}\bigg)^{2} \sin 3\varphi_{0}(t),
\label{eq:Is_phi_0}
\end{eqnarray}
After integrating equation (\ref{phidot_0}) with respect to time, we have
\begin{eqnarray}
\varphi_{0}(t) &=&\varphi_{0}(0)+I_{0}t - \frac{A}{\Omega} \cos \Omega t.
\label{phi_0}
\end{eqnarray}
Inserting equation (\ref{phi_0}) into equation (\ref{eq:Is_phi_0}) and after some algebra \cite{supplement}, we come to:
\begin{eqnarray}
I^{(0)}_{s} &=& Im\bigg\{\sum_{n=-\infty}^{\infty} i^{n} \bigg[ J_{n}  \bigg(\frac{A}{\Omega}\bigg) \nonumber \\ &\times&\bigg[\bigg (\frac{r \tilde{\gamma}_{2}}{2D}\bigg)^{2}-1\bigg]e^{i((n \Omega -I_{0})t-\varphi_{0}(0))} \nonumber \\ &+& J_{n} \bigg(\frac{2A}{\Omega}\bigg)\frac{r \tilde{\gamma}_{2}}{2D} e^{i((n \Omega -2I_{0})t-2\varphi_{0}(0))} \nonumber \\ &-& J_{n} \bigg(\frac{3A}{\Omega}\bigg) \bigg (\frac{r \tilde{\gamma}_{2}}{2D}\bigg)^{2} e^{i((n \Omega -3I_{0})t-3\varphi_{0}(0)} \bigg]\bigg\}.
\label{eq:Is_phi_0_4}
\end{eqnarray}
Shapiro step within this order appears when the ac component of the supercurrent vanishes. This means we have three possible cases; $I_{0}=n\Omega$, $I_{0}=n\Omega/2$, and $I_{0}=n\Omega/3$. Furthermore, in the supplement we show that the $1^{st}$ order leads to fractional subharmonic steps which occur at different set of integers. These locking conditions appear only in the presence of spin-orbit coupling.
\begin{figure}[!ht]
	\includegraphics[width=\linewidth]{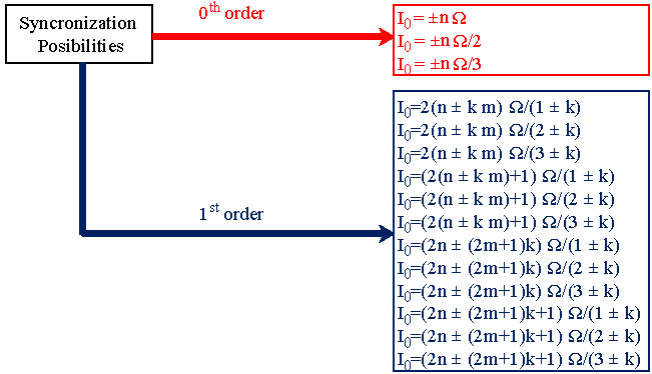}
	\caption{  Condition for Shapiro steps in $\varphi_{0}$-junction, $I_{0}$ is the applied current, $n$ and $m$ take any integer value while $k$ takes the value of $1,2,3$.  \label{4}}	
\end{figure}

So, we find the complete frequency locking conditions for subharmonic steps, which are shown in figure  \ref{4}. In the next section, we demonstrate these conditions for some steps.

\subsection{ DS structure in the IV-characteristics of $\varphi_{0}$ junction}
 \begin{figure}[H]
	\centering
	\includegraphics[width=0.6\linewidth, angle=0]{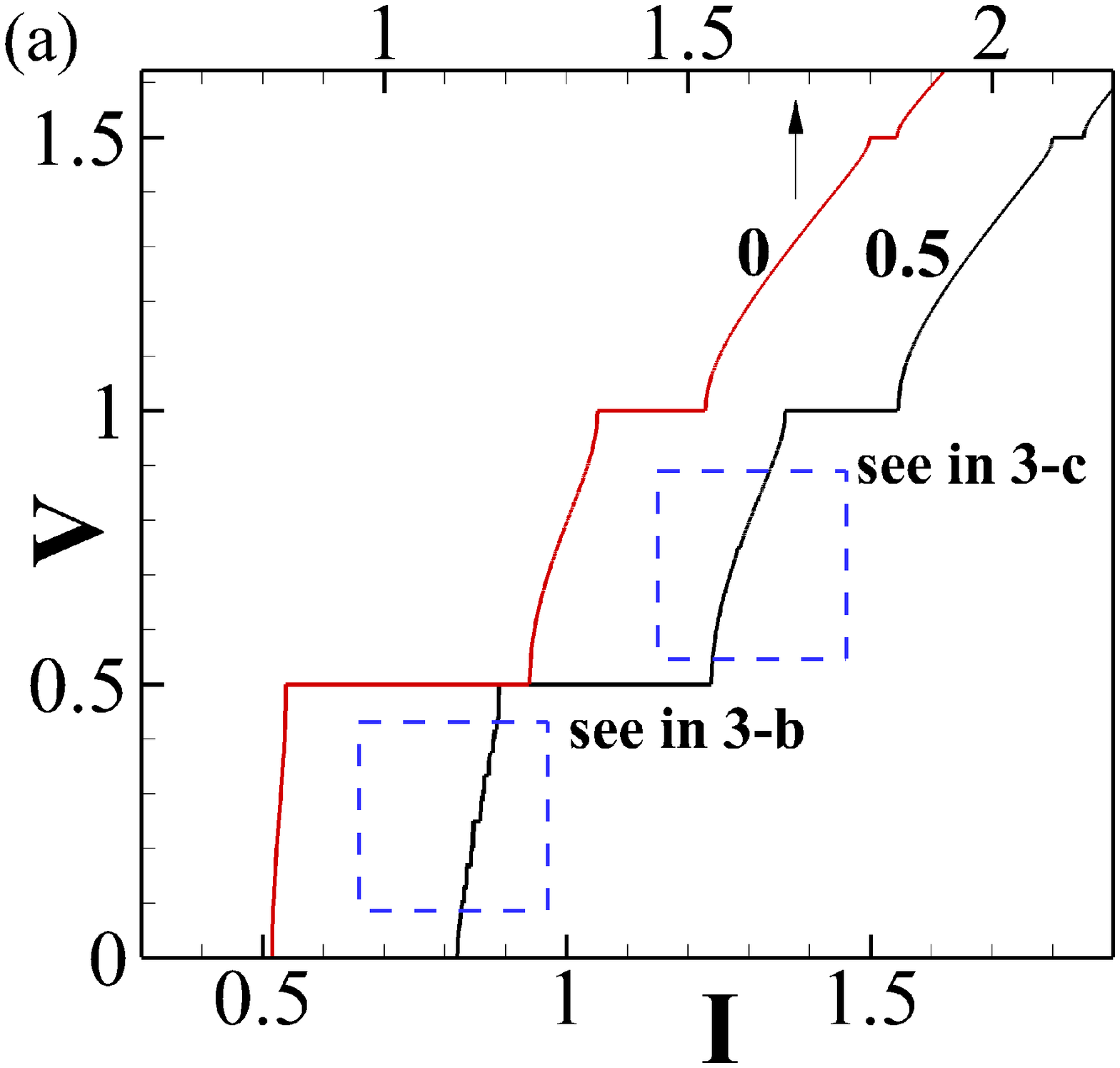}
	\includegraphics[width=0.6\linewidth, angle=0]{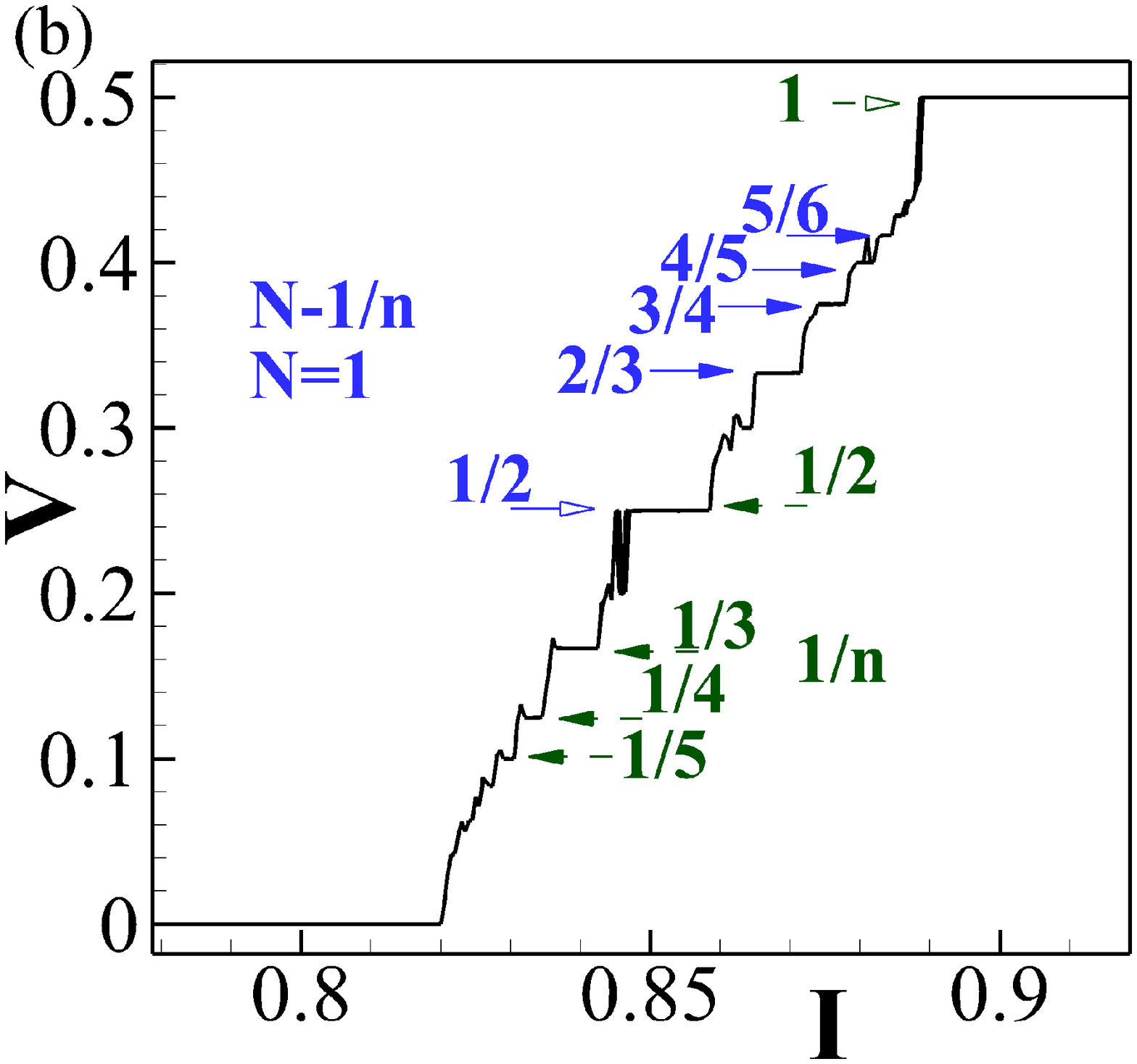}
	\includegraphics[width=0.6\linewidth, angle=0]{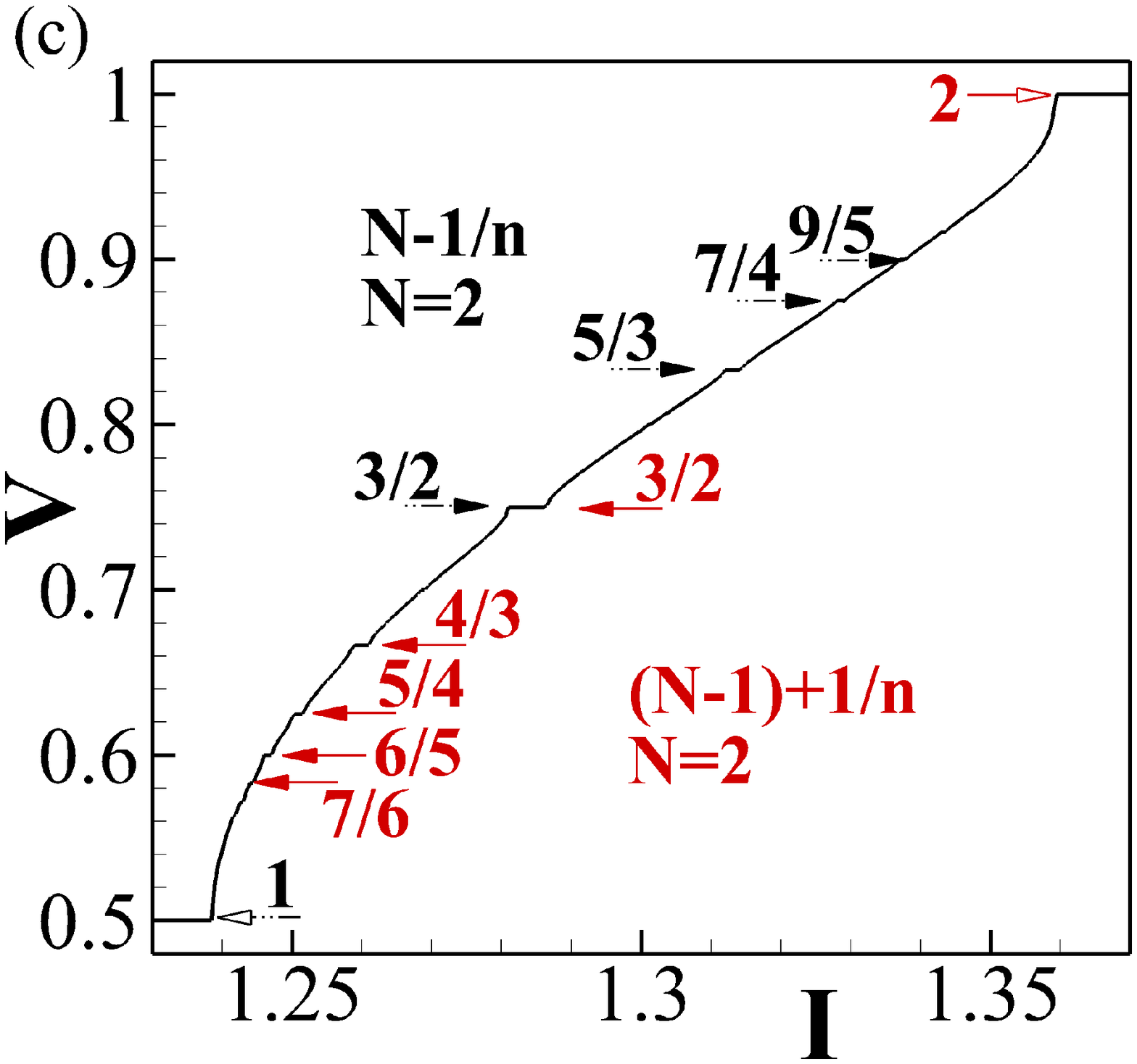}
	\caption{ (a) IV-characteristic  of overdamped $\varphi_{0}$ junction without ($r=0$) and with spin-orbit coupling ($r=0.5$).  Enlarged parts of IV-characteristic marked by rectangles in (a) are shown in (b) and (c). For all figures we take $G=0.2$, $\alpha=0.1$, $A=0.5$ and $\Omega=\Omega_{F}=0.5$. }
	\label{6}
\end{figure}
Here we show the DS structure in the IV-characteristics of $\varphi_{0}$ junction under external electromagnetic radiation with $A=0.5$ and $\Omega=0.5$ and prove their correspondence to the continued fraction formula (\ref{eq:conti}).  Figure \ref{6}(a) demonstrates two IV-characteristics without and with spin-orbit coupling. At $r=0$ the IV-characteristic shows only harmonic Shapiro steps at $V=n\Omega$ with $n$ integer. However, the additional fractional subharmonics appear at $r=0.5$  between the harmonic steps as a result of spin-orbit coupling. Figures ~\ref{6}(b) and \ref{6}(c) demonstrate the enlarged parts of the IV-characteristic shown in figure~\ref{6}(a). We see in figure~\ref{6}(b) the fractional steps between $V=0$ and $V=0.5$ which can be described by the continued fractions of second level $(N-1)+1/n$  and $N-1/n$ with $N=1$ in both cases. In figure~\ref{6}(c) we see the manifestation of second level continued fractions $N-1/n$ and $(N-1)+1/n$ with $N=2$  between voltage steps $V=0.5$ and $V=1$.

\subsubsection{Effect of $r$ on the width of the subharmonic steps }
\begin{figure}[H]
	\centering
	\includegraphics[width=0.6\linewidth, angle=0]{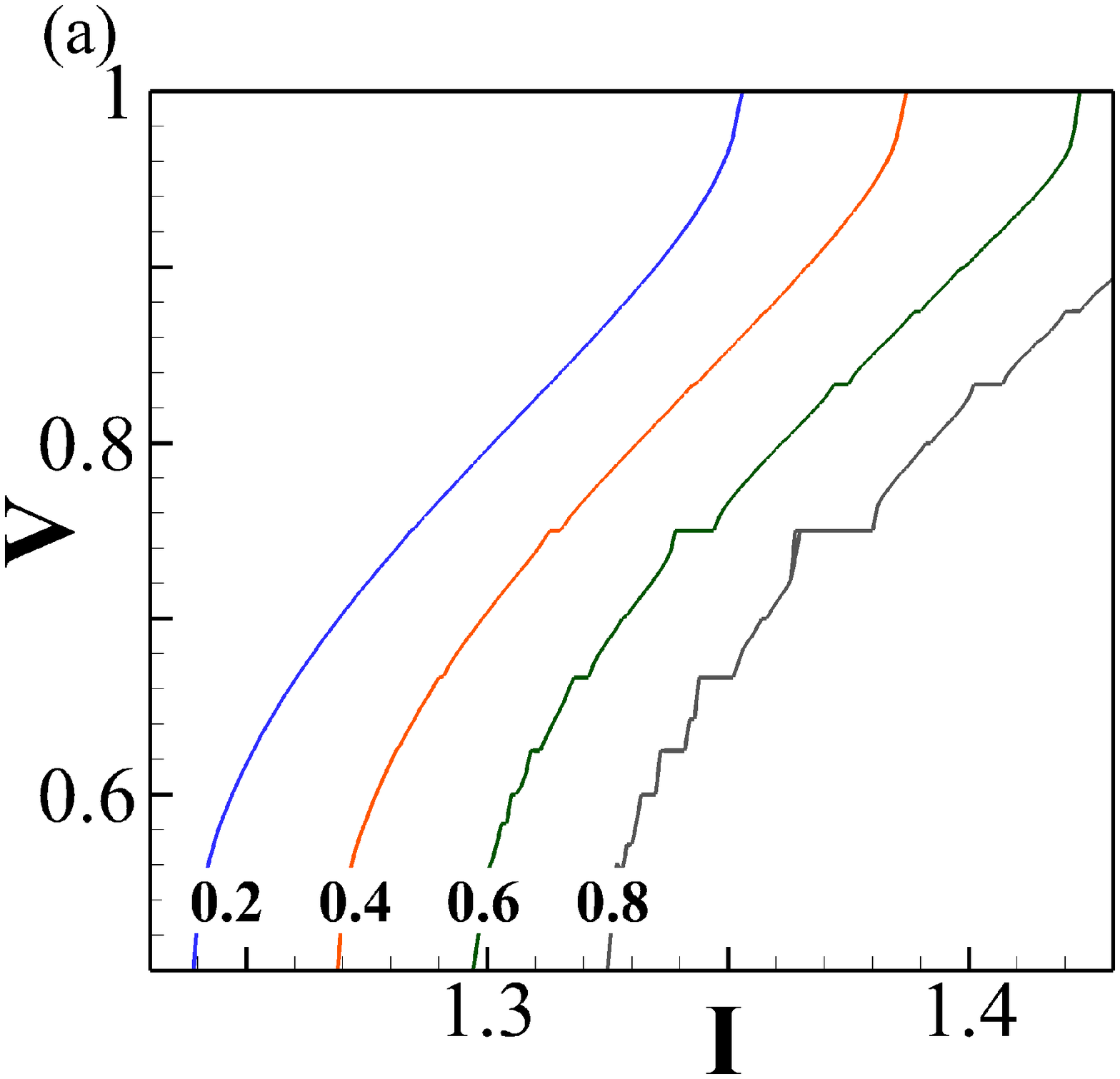}
	\includegraphics[width=0.6\linewidth, angle=0]{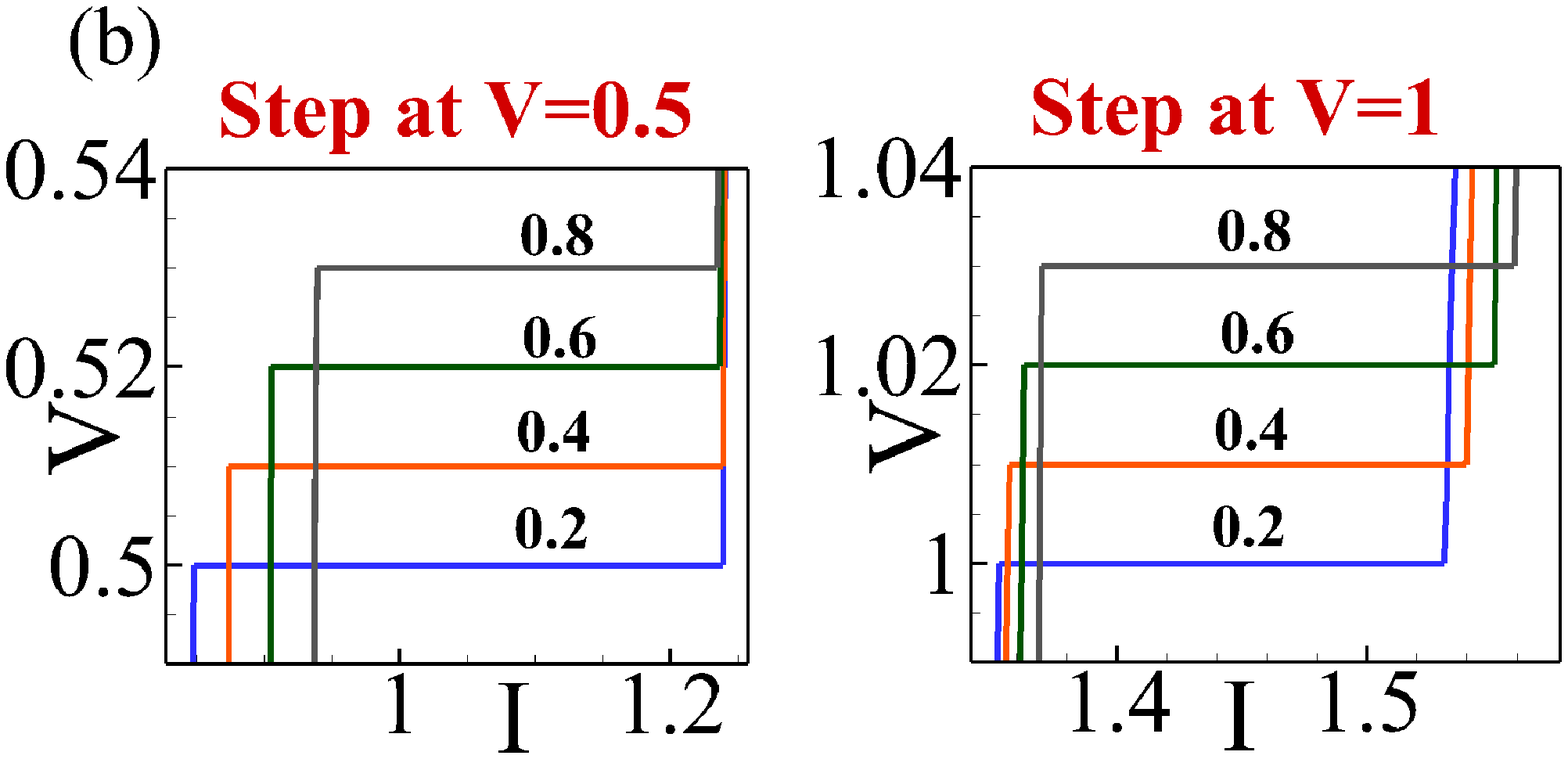}
	\includegraphics[width=0.6\linewidth, angle=0]{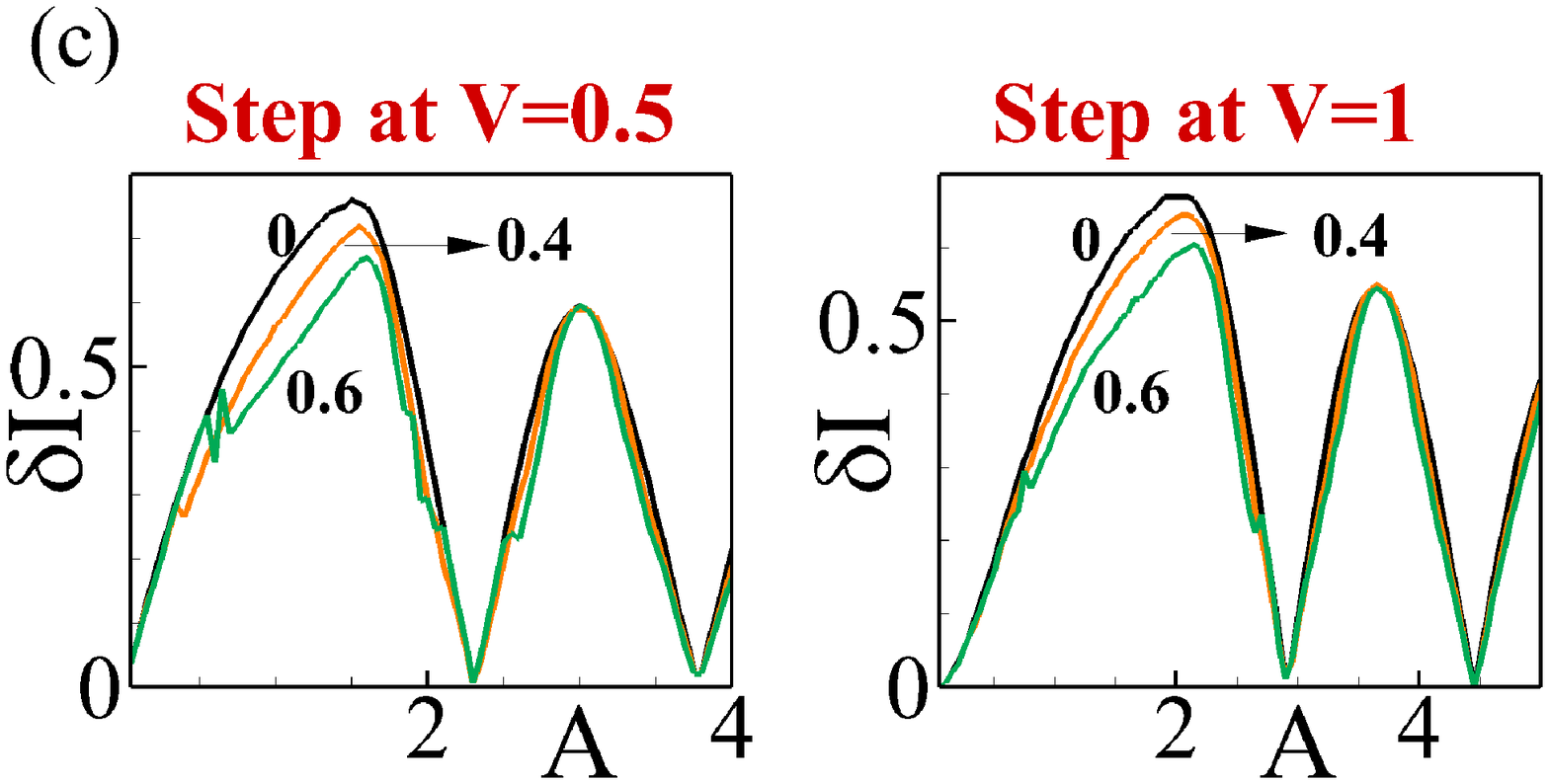}
	\includegraphics[width=0.6\linewidth, angle=0]{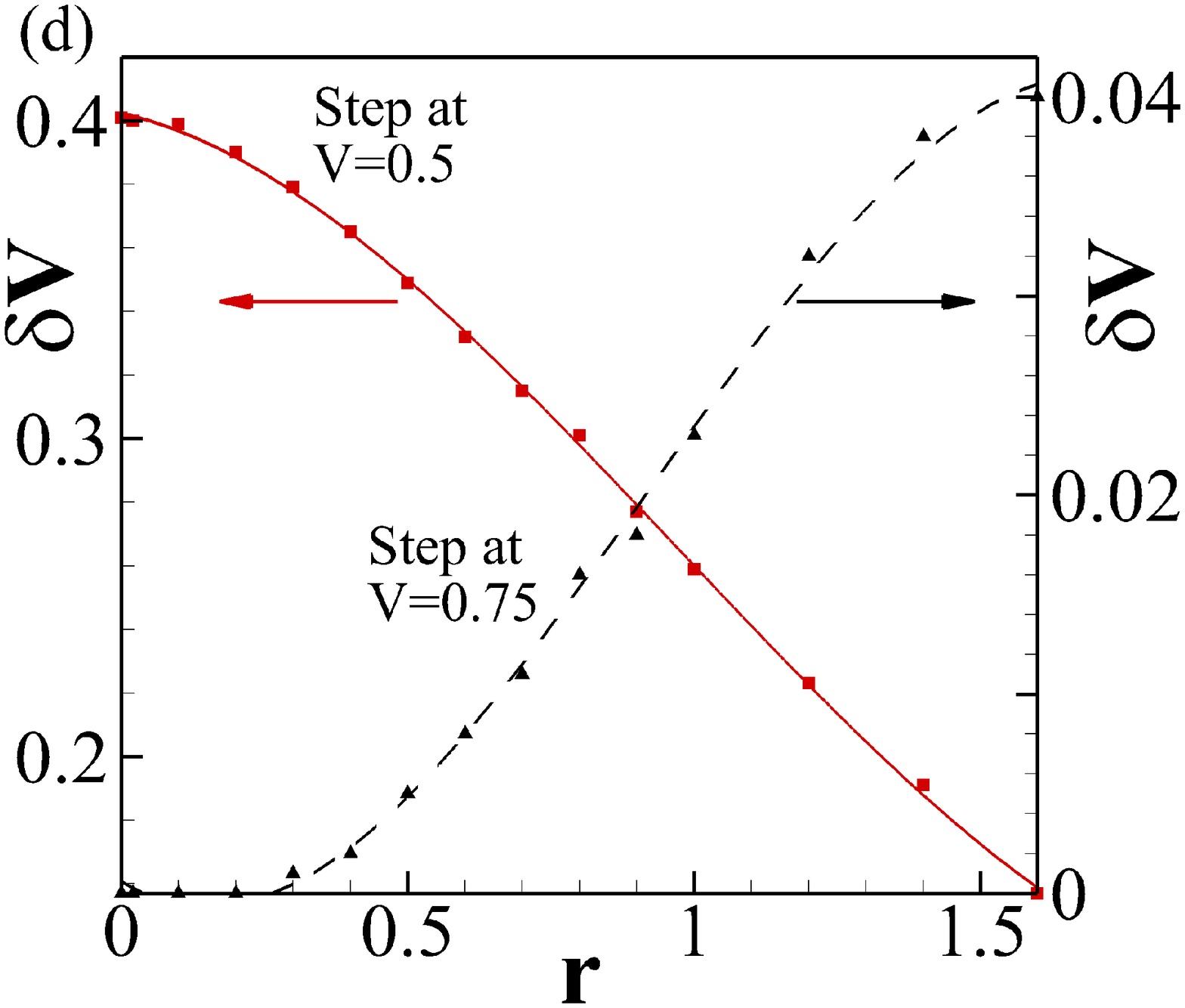}
	\caption{ (a) Parts of the IV-characteristics at $\Omega=0.5$, $G=0.2$ and different values of $r$. The curves at $r=0.4$ , $r=0.6$ and $r=0.8$ have been shifted for clarity by $\Delta I= 0.03$, $=0.06$, and $0.09$, respectively, relative to the IV-characteristic at $r=0.2$;  (b) Harmonic Shapiro steps at $V=0.5$ and $V=1$ at different value of $r$. The curves at $r=0.4$ , $0.6$ and $0.8$ shifted up by $\Delta V= 0.01$, $\Delta V=0.02$, and $\Delta V=0.03$, respectively, relative to the IV-characteristic at $r=0.2$; (c) Width of the harmonic Shapiro steps at $V=\Omega=0.5$ and $V=2\Omega=1$ as a function of $A$ at different $r$; (d) Width's r-dependence for the harmonic step at $V=0.5$  and subharmonic step  at $V=0.75$.  \label{5}}	
\end{figure}
Let us now inspect the effect of spin-orbit coupling on the presence of subharmonic steps in details. Figure~\ref{5}(a) compares the IV-characteristics at different value of $r$ and we see that the enhanced subharmonic steps appear at large value of $r$.

In figure~\ref{5}(b) we compare the harmonic Shapiro steps at $V=\Omega$ and $V=2\Omega$ at different value of $r$. We find qualitatively different behavior:  the width of the first harmonic Shapiro step at $V=\Omega$ decreases slightly with increasing $r$, while the Shapiro step at $V=2\Omega$ demonstrates some horizontal shift with almost the same width for different values of $r$.

The corresponding width dependence  of these steps as a function of amplitude of external radiation $A$ is shown  figure~\ref{5}(c).  For the given simulation parameters, we observe a significant change of the Bessel dependence of the Shapiro step width with increase in $r$ in the range $A\sim0.5 -2$.

However, as we notice above  in figure~\ref{5}(a), the width of fractional Shapiro steps increases with  $r$. The $r$-dependence of the harmonic  step width at $V=0.5$  and subharmonic step at $V=0.75$ is shown in figure~\ref{5}(d). We see the qualitatively different behavior: harmonic  step width is decreased, while a reversal dependence occurs for subharmonic step.

The perterbative analysis shows that the width of all steps, harmonic and subharmonic, is not Bessel function of $r$ \cite{supplement}, moreover, we have found that the width of subharmonic steps is proportional to $B_{x,y}$=$J_{x}(k A/\Omega)J_{y}(A/\Omega)$, where $k=1,2,3$.

\subsubsection{Effect of $G$ on the appearance of the DS structure}

Another important parameter which can control the appearance of the subharmonic steps is a ratio of Josephson and magnetic anisotropy energies $G=E_{J}/Kv$. In Ref.\cite{Shukrinovepls2018} it is shown that a reorientation of easy axis from  $m_{z}=1$ to $m_{y}=1$ occurs  in the regime of small Josephson frequency and large  $G>20$.  Here we show that the subharmonic steps are enhanced in the regime before a complete reorientation occurs.

\begin{figure}[H]
	\centering
	\includegraphics[width=0.7\linewidth, angle=0]{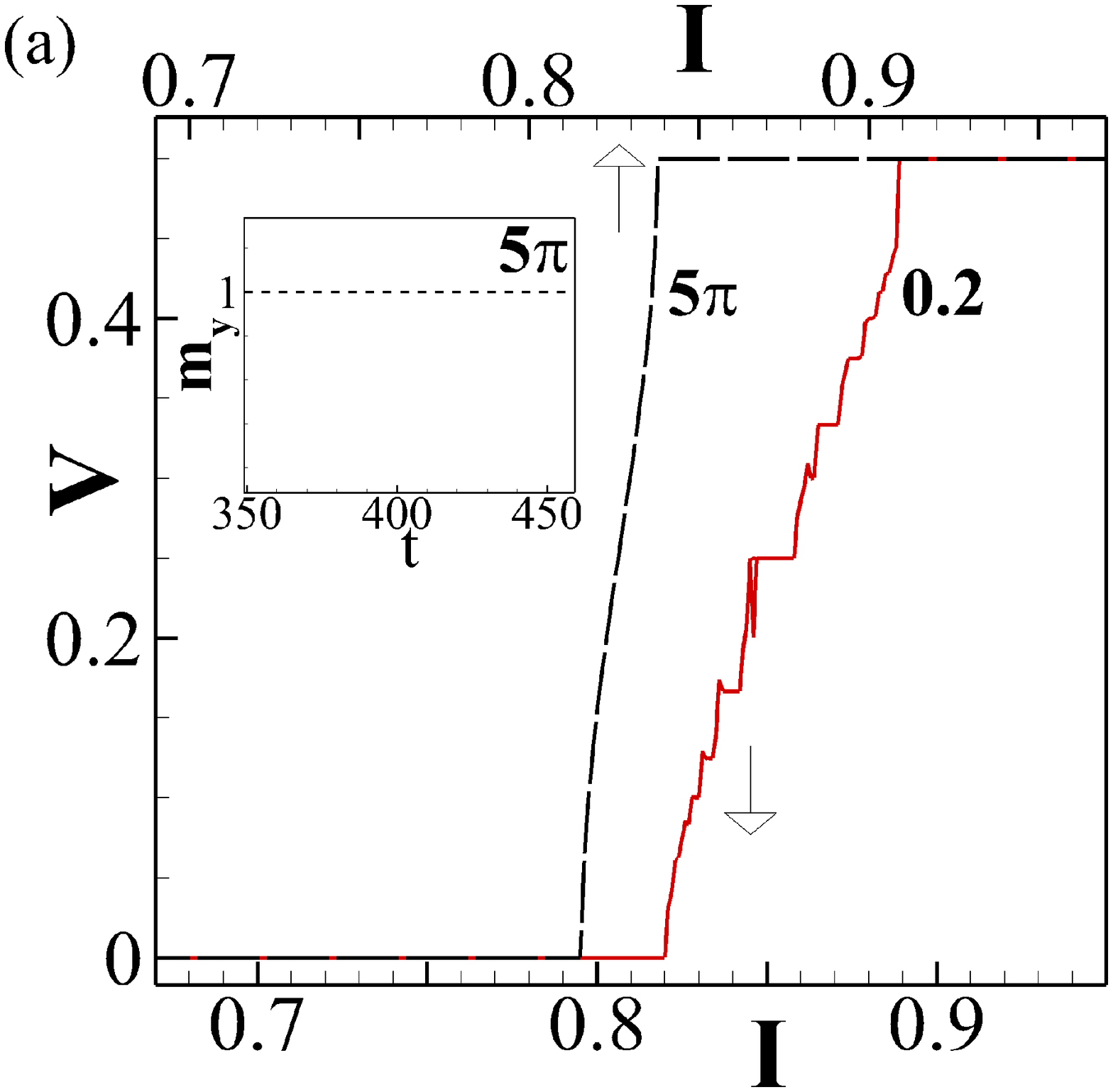}
	\includegraphics[width=0.7\linewidth, angle=0]{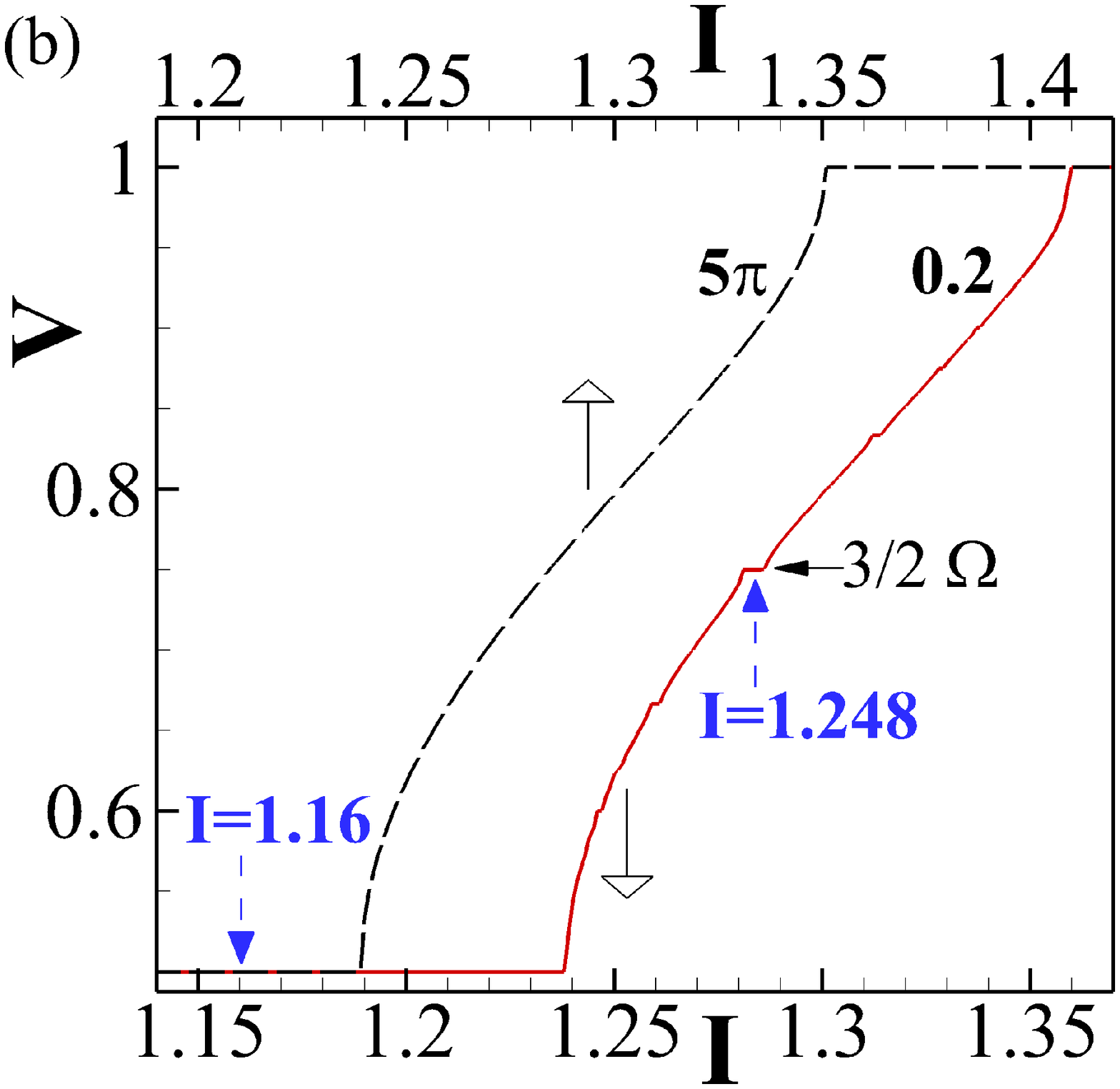}
	\includegraphics[height=0.42\linewidth, angle=0]{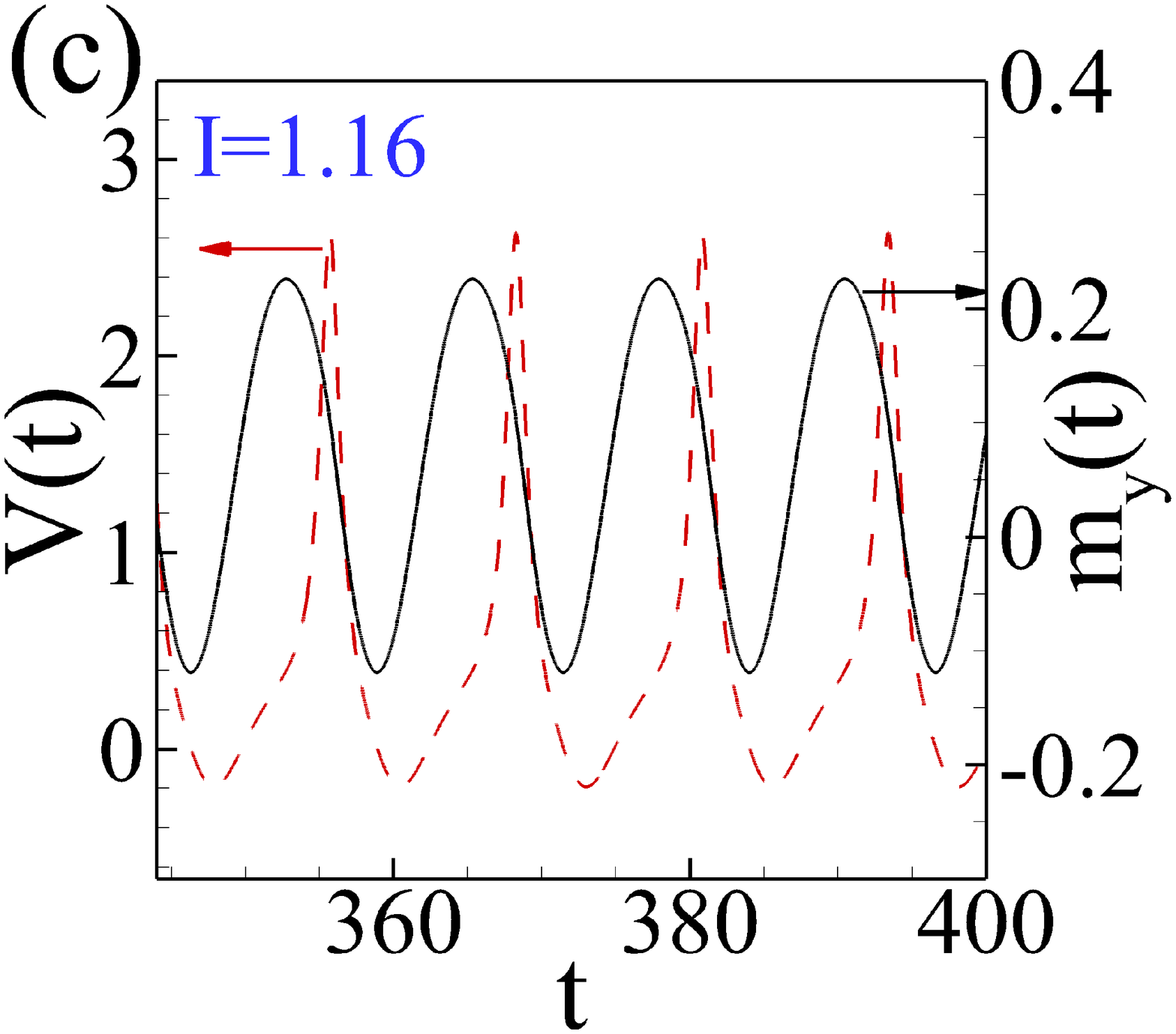}
	\includegraphics[width=0.45\linewidth, angle=0]{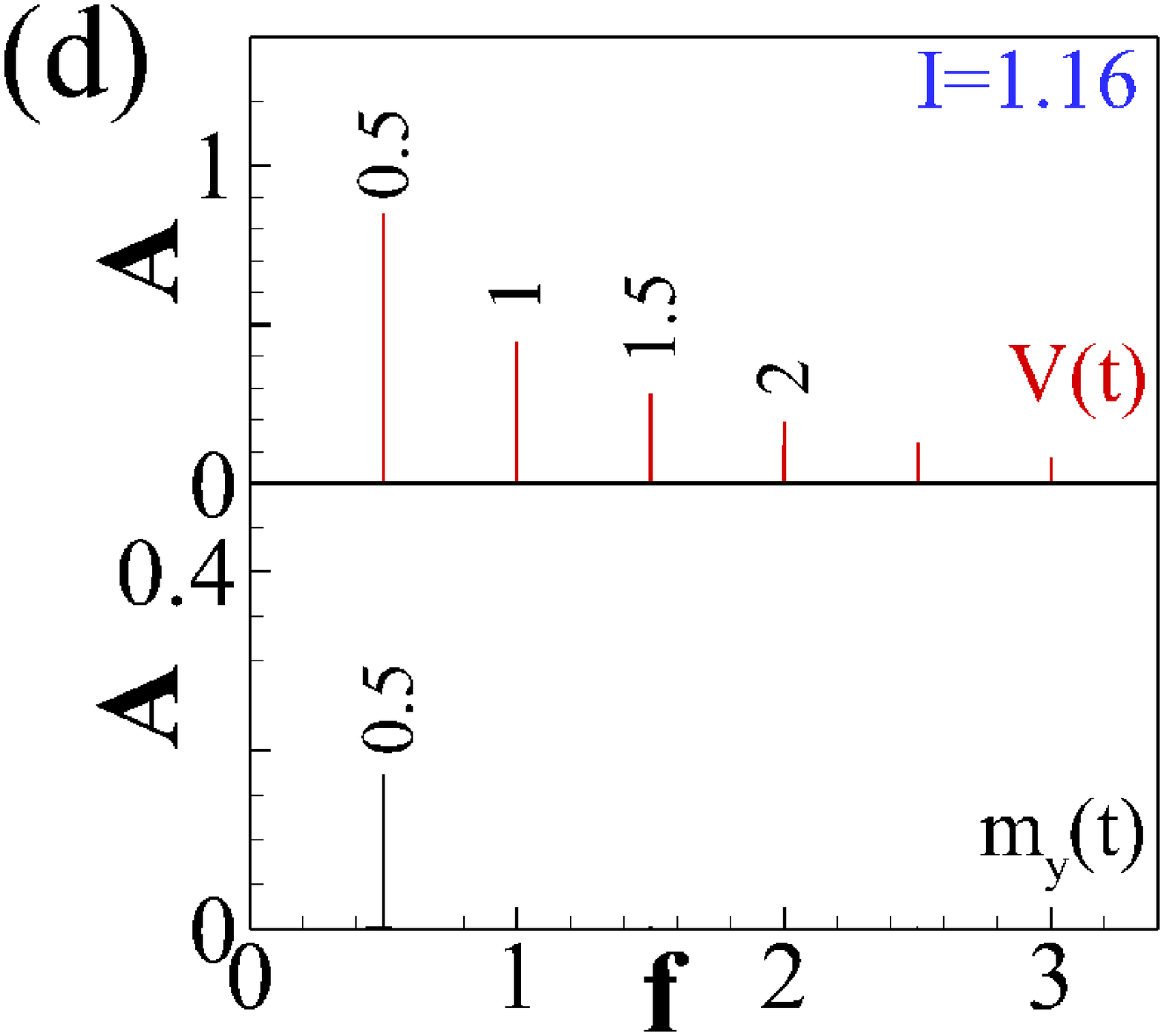}
	\includegraphics[height=0.42\linewidth, angle=0]{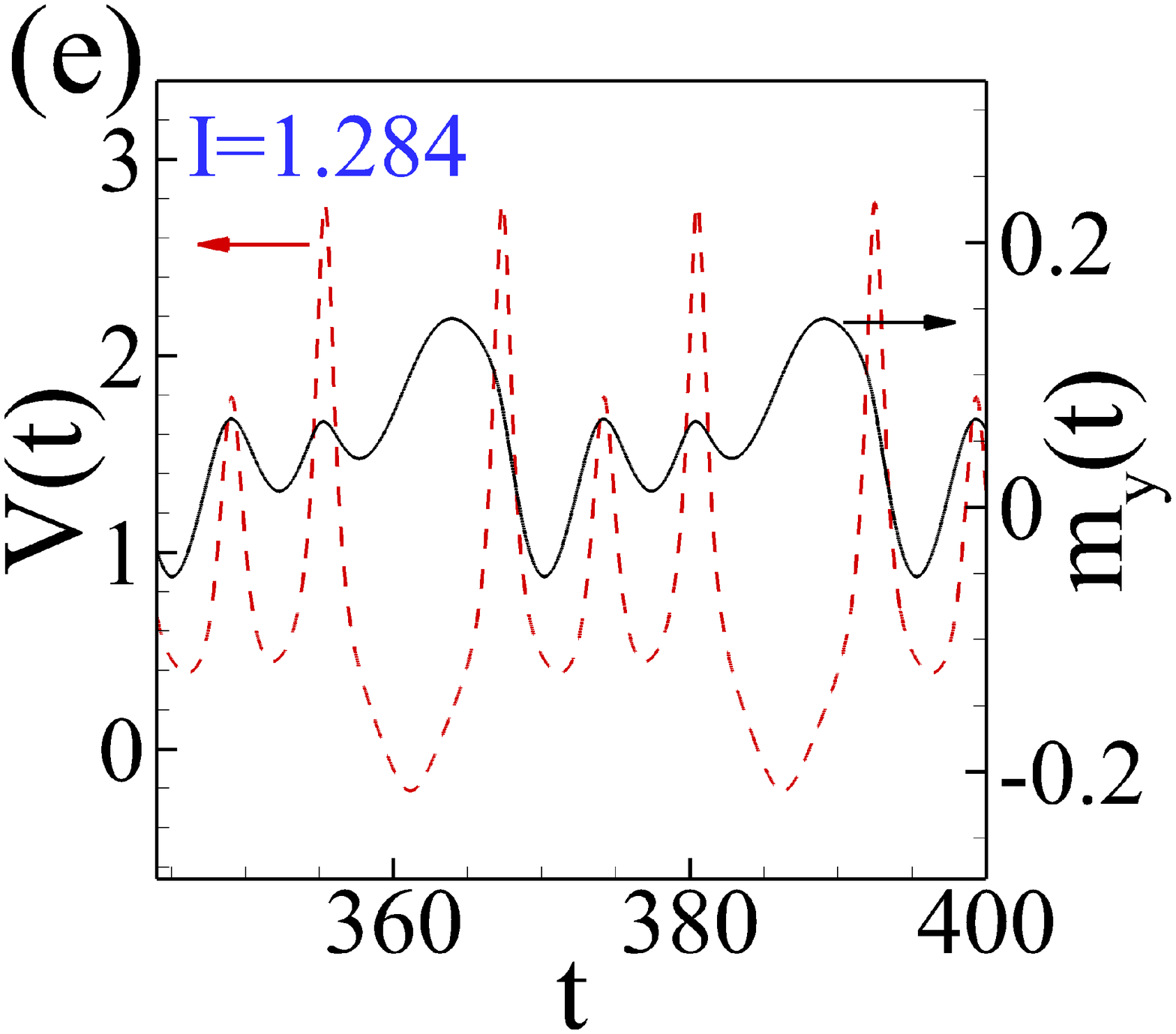}
	\includegraphics[width=0.45\linewidth, angle=0]{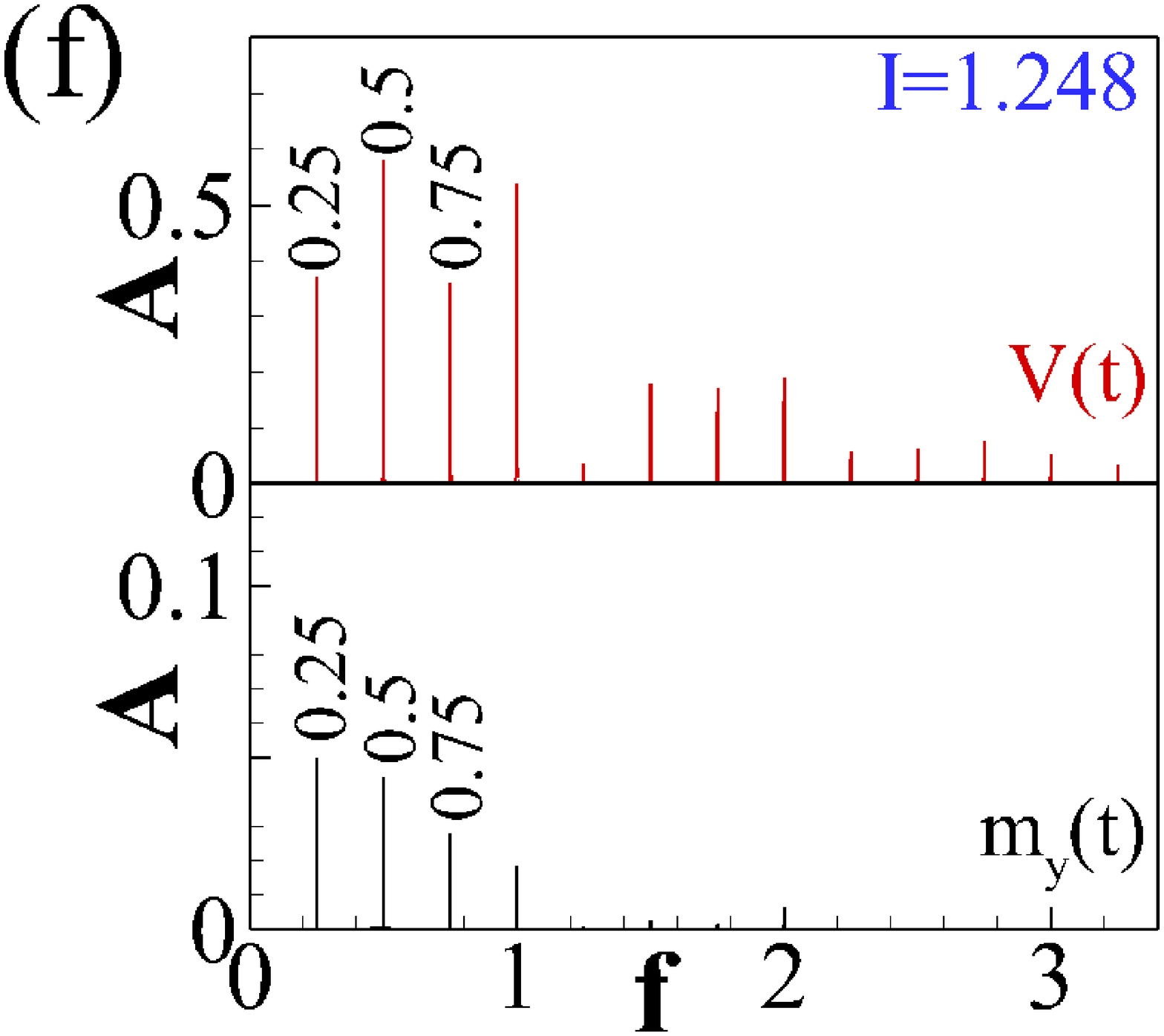}
	\caption{(a) Enlarged parts of the IV-characteristics in the interval $0<V<0.5$ at $G=0.2$ and $G=5\pi$. The inset shows the temporal dependence of $m_{y}(t)$ at $G=5\pi$; (b) The same in the interval $0.5<V<1$; (c) Temporal dependence of $V(t)$ and $m_{y}(t)$ at $I=1.16$, $G=0.2$ and in (d) The corresponding results of the FFT analysis; (e) Temporal dependence of $V(t)$ and $m_{y}(t)$ at $I=1.284$, $G=0.2$ and in (f) The corresponding results of the FFT analysis. For all figures we take $\alpha=0.1$, $\Omega_{F}=0.5$, and $r=0.5$. \label{8}}	
\end{figure}

In figure \ref{8}(a) and (b)  we show the enlarged parts of the IV-characteristics at $G=0.2$ and $G=5\pi$. At $G=0.2$ the subharmonic steps appear mostly between $V=0$ and $V=0.5$ [see figure \ref{8}(a)], while much less number of steps with smaller width appear between $V=0.5$ and $V=1$ [see figure \ref{8}(b)].  All these fractional steps disappear at $G=5\pi$ when a reorientation of easy axis occurs to $m_{y}(t)=1$ (see insets).

Figure \ref{8}(c) shows the temporal dependence of $V(t)$ and $m_{y}(t)$ at $I=1.16$ (step at $V=\Omega$). The temporal dependence for both $V(t)$ and $m_{y}(t)$ is regular. Results of the corresponding Fast Fourier Transform (FFT) analysis  for $V(t)$ and $m_{y}(t)$ are shown in figure \ref{8}(d). From the FFT analysis it is clear that the oscillation frequency of $m_{y}(t)$ is locked to the external frequency $\Omega=0.5$ as well as to $\Omega_{J}=0.5$. In addition to this, the FFT analysis for $V(t)$ show harmonics of $n\Omega$ with integer $n$.

Next, we have recorded the temporal dependence of $V(t)$ and $m_{y}(t)$  at $I=1.284$ for fractional step at $V=0.75$, presented in figure \ref{8}(e) and the corresponding  result of the FFT analysis is shown in figure \ref{8}(f). Here the temporal dependence for both $V(t)$ and $m_{y}(t)$ is more complex. The oscillation amplitude for $m_{y}(t)$ is smaller than in case $V=0.5$ [see figure \ref{8}(c)]. The FFT analysis show that a locking between the precession frequency for $m_{y}$ and Josephson frequency occurs at fractions of $\Omega$.

Based on the presented results, we can come to the following conclusions. The first is locking of the magnetic moment precession  at $I=1.16$ and $G=0.5$ to the external radiation with frequency $\Omega=0.5$. At $I=1.284$ we also observe the locking of magnetic moment precession to the subharmonic frequency $3\Omega/2$ which correspond to the step at $V=0.75$. The appearance of Shapiro steps at $V=n\Omega$ is not related to the magnetic moment precession. Shapiro step at $V=0.5$ appears at $G=5\pi$ with $m_{y}(t)=1$ when the precession frequency is zero.

Finally, we present some estimations for the experimental realization based on data in Ref's.\cite{buzdin2009,Assouline2019,exper_data1,exper_data2,exper_data3,exper_data4}. The main parameter which controls the appearance of the current steps is $G =E_{J}/(K v)$ which represents the ratio of the Josephson and the magnetic anisotropy energy. With flux quantum $\Phi_{0}=2.067833\times10^{-15} (J/A)$, critical current $I_{c}=4 \mu A$, volume of the ferromagnet $v=d\times l\times w\sim 150 nm \times 2\mu m \times 2\mu m $, and anisotropy constant $K=500$ ($J /m^{3}$), we obtain $G=\Phi_{0} I_{c}/(K v)=0.4$. Actually, for stronger anisotropy, parameter G is smaller. Play with the value of anisotropy parameter, we may have G as much less or much large than one.

\section{Conclusion}

In this work we solve the dynamical equations for $\varphi_{0}$ junction which describe the coupling between Josephson phase and magnetic moment through spin-orbit coupling. Using the high-frequency perterbative method we predict the appearance of subharmonic steps in the presence of spin-orbit coupling. Furthermore we confirm these prediction using the exact numerical scheme. The IV-characteristic of the $\varphi_{0}$ junction demonstrates devil's staircase structures of Shapiro steps. The origin of the found steps related to the  synchronization between Josephson and magnetic moment oscillations. The position of the found steps follow continued fraction formula. The structure and width of these steps depends on the ratio of Josephson and magnetic energies and spin-orbit coupling. Our calculations predict that spin-orbit coupling manifests itself through the appearance of subharmonic steps in the IV-characteristics, thus providing new insight into the precise nature of the current-phase relation and new
opportunities for potential applications.

\section{Acknowledgments}

The authors thank I. Rahmonov and A. Mazanik for fruitful discussions. The reported study was partially funded by the RFBR
research projects  18-02-00318, 18-52-45011-IND. Numerical calculations have been made in the framework of the RSF project 18-71-10095.

\newpage
\clearpage
\newpage
\newpage
\setcounter{figure}{0}
\setcounter{section}{0}
\setcounter{equation}{0}
\begin{widetext}
	\section*{	Supplemental Material to ``Devil's staircase structure in $\varphi_{0}$ junction''}
\section{Linearized Landau-Lifshitz-Gilbert equation}
If the deviation of the magnetic moment from the equilibrium point due to Josephson energy is small( $G<1$), we can linearized the Landau-Lifshitz-Gilbert (LLG) equation.  In its general form the LLG equation reads
\begin{eqnarray}
\dfrac{d\bm{M}}{dt}=-\gamma\bm{M}\times\bm{H}_e+\dfrac{\alpha}{\Arrowvert\bm{M}\Arrowvert}\left(\bm{M}\times \frac{d\bm{M}}{dt} \right),  \label{eq:LLGa}
\end{eqnarray}
where $\alpha$ is the Gilbert damping and $\gamma$ is the gyromagnetic ratio.  We assume that the effective magnetic field and magnetization can be written as sums of constant and alternating parts
\begin{equation}
\bm{H}_{e}=\bm{H}_{0}+\bm{\tilde{H}}, \ \ \  \ \ \bm{M}=\bm{M}_{s} + \bm{\tilde{M}},
\label{eq:stead_alter}
\end{equation}
where the components of $\bm{H}_{0}$ are $(0,0,H_{0})$  with $H_{0}=K_{an}/M_{s}=\omega_{F}/\gamma$,  $K_{an}$ is the  magnetic anisotropy constant, $M_{s}=\Arrowvert\textbf{M}\Arrowvert$ is considered as a constant and equal to the saturation value of the magnetization, $\omega_{F}$ is the ferromagnetic resonance frequency and $\gamma$ is the gyromagnetic ratio. The components of $\bm{\tilde{H}}$ are  $(\tilde{H}_{x}, \tilde{H}_{y},0)$, the components of $\bm{M}_{s}$ are $ (0,0,M_{z})$ and those of $\bm{\tilde{M}}$ are $(\tilde{M}_{x}, \tilde{M}_{y}, 0)$. The magnitude of alternating parts are considered smaller than the steady parts, i.e.  $\tilde{H} << H_{a} $, $\tilde{M}<< M_{z}$. In what follow we use the dimensionless formula. We normalize $H_{e}$ to $H_{0}$, $M_{i}$ ($i=x,y$) is normalized to $M_{s}$, time is normalized to $\omega_{c}^{-1}$ ($t\longrightarrow t\omega_{c}^{-1}$) where $\omega_{c}$ is the characteristic frequency of JJ, and $\omega_{F}$ is normalized to $\omega_{c}$. The linearization of (\ref{eq:LLGa}) can be found by inserting (\ref{eq:stead_alter}) into (\ref{eq:LLGa}) and neglecting the products of the alternating parts. In the dimensionless form the linearized LLG is then reads as 
\begin{eqnarray}
\frac{d\bm{\tilde{m}}}{dt} + \Omega_{F} \bm{\tilde{m}} \times \bm{h}_{0}+\alpha \left(\frac{d\bm{\tilde{m}}}{dt}\times \bm{m}_{s}\right)= - \Omega_{F}  \bm{m}_{s}\times\bm{\tilde{h}},
\label{eq:LinLLG3}
\end{eqnarray}
where the components of $\bm{h}_{0}$ are $(0,0,1)$, $\bm{\tilde{h}}$ are  $(h_{x}, h_{y},0)$, $\bm{m}_{s}$ are $(0,0,1)$, and $\bm{\tilde{m}}$ are  $(m_{x}, m_{y},0)$.	We assume a harmonic time dependence for the effective field $\bm{\tilde{h}}$ and $\bm{\tilde{m}}$ in the form of $\bm{\tilde{m}}= \textit{\textbf{m}} \ e^{i\Omega_{p} t}$ and $\bm{\tilde{h}} = \textit{\textbf{h}} \ e^{i\Omega_{p} t}$, where $\Omega_{p}$ is the precession frequency normalized to $\omega_{c}$ ($\Omega_{p}=\omega_{p}/\omega_{c}$). According to this, the linearized LLG reads as
\begin{equation}
i \Omega_{p} \textit{\textbf{m}} +\Omega_{F}  \textit{\textbf{m}} \times \bm{h}_{0}+i \Omega_{p} \alpha  \textit{\textbf{m}}\times \bm{m}_{s}= -\Omega_{F}  \bm{m}_{s}\times\textit{\textbf{h}}.
\label{eq:LinLLG2}
\end{equation}
Projecting Eq.~(\ref{eq:LinLLG2}) on the axes of Cartesian coordinate system we find the real part for $m_{y}(t)$ (we will take $\tilde{m}_{y}\equiv m_{y}(t)$, $\tilde{h}_{i}\equiv h_{i}(t)$ and $i=x,y$.)
\begin{eqnarray}
Re\{m_{y}(t)\} &=& \left[ \frac{-\gamma_{1} h_{x}(t)+ \gamma_{2}  h_{y}(t)}{\left( 1-(1+\alpha^{2})\frac{\Omega_{p}^{2}}{\Omega_{F}^{2}}\right) ^{2} + 4 \alpha^{2}\frac{\Omega_{p}^{2}}{\Omega_{F}^{2}}}\right].
\label{eq:m_expre}
\end{eqnarray}
where $\gamma_{1}=2  \alpha \frac{\Omega_{p}^{2}}{\Omega_{F}^{2}}$ and $\gamma_{2}=\left( 1-(1-\alpha^{2}) \frac{\Omega_{p}^{2}}{\Omega_{F}^{2}}\right)$. To find $m_{y}(t)$, we need the x and y-component of the effective field. Here we have $h_{x}=0$ and $h_{y}=Gr\sin(\varphi-rm_{y})$. Since $m_{y}<<1$, we can write $\sin(\varphi(t)-rm_{y}(t))\approx \sin \varphi(t) - r m_{y}(t) \cos \varphi(t)$. Using Eq.(\ref{eq:m_expre}), the expression of $m_{y}(t)$  is given by
\begin{eqnarray}
m_{y}(t)=\frac{\tilde{\gamma}_{2}\sin \varphi(t)}{D+r\tilde{\gamma}_{2} \cos\varphi(t)},
\label{eqa_my1}
\end{eqnarray}
where $\tilde{\gamma}_{2}=Gr\gamma_{2}$, and $D=\left( 1-(1+\alpha^{2})\frac{\Omega_{p}^{2}}{\Omega_{F}^{2}}\right) ^{2} + 4 \alpha^{2}\frac{\Omega_{p}^{2}}{\Omega_{F}^{2}}$. After using the expansion $1/(1+x)\approx 1-x+..$, we come to
\begin{eqnarray}
m_{y}(t)&\approx&\frac{\tilde{\gamma}_{2}}{D}\sin \varphi(t)-\frac{r\tilde{\gamma}^{2}_{2}}{2D^{2}} \sin 2\varphi(t),
\label{eqa_msy}
\end{eqnarray}

\section{Origin of subharmonic step in $\varphi_{0}$ junction}

To find the origin of subharmonic step, we use high-frequency limit \cite{Kornev}: 
\begin{equation}
\Omega>>1 , \ \  \beta_{c} \Omega >>1, \ \   A>>1
\end{equation}
where $\Omega$ is the frequency of the external electromagnetic radiation normalized to $\omega_{c}$ (characteristic frequency of JJ),  $\beta_{c}$ is the McCumber parameter, and $A$ is the amplitude of the external electromagnetic radiation normalized to $I_{ac}$.

Using Eq.(\ref{eqa_msy}), the RSJ equation reads as
\begin{eqnarray}
\dot{\varphi}(t)&=&I + A \sin \Omega t + \bigg[\left( \frac{r \tilde{\gamma}_{2}}{2D}\right)^{2}  -1\bigg]\sin \varphi(t)+\frac{r \tilde{\gamma}_{2}}{2D} \bigg( \sin 2\varphi(t) - \frac{r \tilde{\gamma}_{2}}{2D} \sin 3\varphi(t)\bigg),
\label{eqa12}
\end{eqnarray}
Expend $I$ and $\varphi$ as \cite{Shukrinov2011}
\begin{eqnarray}
\varphi(t) = \sum_{n}^{} \epsilon^{n} \varphi_{n}(t), \ \ \  I = \sum_{n=0}^{\infty} \epsilon^{n} I_{n}.
\label{series2}
\end{eqnarray}
where $I_{0}$ is the bias current, $\epsilon<<1$ and $I_{n}$ for $n > 0$ are determined from the condition of the absence of additional dc voltage: $\lim_{T\longrightarrow\infty}$ $\int_{0}^{T}$ $\dot{\varphi}_{n} dt=0$\cite{Likharev}. Using (\ref{series2}), the equations for $\dot{\varphi}_{n}$ can be obtained by equating terms in the same order of $\epsilon$. Then, the expression of $\dot{\varphi}_{n}$ is given by 
\begin{eqnarray}
\dot{\varphi}_{n}(t) =I_{n} +f_{n} (t),
\label{maineeq}
\end{eqnarray}

The second and third terms in Eq.(\ref{eqa12}) after using $\varphi(t)=\varphi_{0}(t)+\epsilon \varphi_{1}(t)$ (to the first order) read as
\begin{eqnarray}
&& \bigg[\left( \frac{r \tilde{\gamma}_{2}}{2D}\right)^{2}  -1\bigg]\sin \varphi(t)+\frac{r \tilde{\gamma}_{2}}{2D} \bigg( \sin 2\varphi(t) - \frac{r \tilde{\gamma}_{2}}{2D} \sin 3\varphi(t)\bigg) \approx \nonumber \\&&
\bigg[\left( \frac{r \tilde{\gamma}_{2}}{2D}\right)^{2}  -1\bigg]\bigg[\sin \varphi_{0}(t) + \epsilon \varphi_{1}(t) \cos \varphi_{0}(t) \bigg]+ \frac{r \tilde{\gamma}_{2}}{2D} \bigg( \sin 2\varphi_{0}(t) +2 \epsilon \varphi_{1}(t) \cos 2\varphi_{0}(t) \nonumber \\ &-& \frac{r \tilde{\gamma}_{2}}{2D} \bigg[\sin 3\varphi_{0}(t) +3 \epsilon \varphi_{1}(t) \cos 3\varphi_{0}(t)\bigg]\bigg).
\label{eqa2}
\end{eqnarray}

Next, we consider 
\begin{eqnarray}
f_{0}(t) &=& A \sin \Omega t, \nonumber \\
f_{1}(t) &=& \bigg[\bigg (\frac{r \tilde{\gamma}_{2}}{2D}\bigg)^{2}-1\bigg]\sin\varphi_{0}(t) +\frac{r \tilde{\gamma}_{2}}{2D} \sin 2\varphi_{0}(t) - \bigg (\frac{r \tilde{\gamma}_{2}}{2D}\bigg)^{2} \sin 3\varphi_{0}(t), \nonumber \\
f_{2}(t) &=& \bigg[\bigg (\frac{r \tilde{\gamma}_{2}}{2D}\bigg)^{2}-1\bigg] \varphi_{1}(t)\cos\varphi_{0}(t)+2\frac{r \tilde{\gamma}_{2}}{2D} \varphi_{1}(t)\cos 2\varphi_{0}(t) - 3\bigg (\frac{r \tilde{\gamma}_{2}}{2D}\bigg)^{2} \varphi_{1}(t)\cos 3\varphi_{0}(t).\nonumber \\
\end{eqnarray}
The zeroth order with $n = 0$ represents the autonomous IV-characteristic of the junction. In this case 
\begin{eqnarray}
\dot{\varphi}_{0}(t) &=&I_{0} + A \sin \Omega t,
\label{phi_0dot}
\end{eqnarray}
and the suppercurrent is given by $I^{(0)}_{s}=-f_{1}(t)$ (see Eq.(\ref{maineeq}))  
\begin{eqnarray}
I^{(0)}_{s} &=& \bigg[1-\bigg (\frac{r \tilde{\gamma}_{2}}{2D}\bigg)^{2}\bigg]\sin\varphi_{0}(t) -\frac{r \tilde{\gamma}_{2}}{2D} \sin 2\varphi_{0}(t) + \bigg (\frac{r \tilde{\gamma}_{2}}{2D}\bigg)^{2} \sin 3\varphi_{0}(t), 
\label{eq:Is_phi_0s}
\end{eqnarray}
After integrating (\ref{phi_0dot}) with respect to time, we come to
\begin{eqnarray}
\varphi_{0}(t) &=&\varphi_{0}(0)+I_{0}t - \frac{A}{\Omega} \cos \Omega t.
\label{eq:phi_0}
\end{eqnarray}
Next, we insert  (\ref{eq:phi_0})  into (\ref{eq:Is_phi_0s}) and use $\sin(-x)=-\sin x$. The suppercurrent is given by
\begin{eqnarray}
I^{(0)}_{s} &=&\bigg[\bigg (\frac{r \tilde{\gamma}_{2}}{2D}\bigg)^{2}-1\bigg]\sin\bigg(\frac{A}{\Omega} \cos \Omega t-I_{0}t-\varphi_{0}(0)\bigg) +\frac{r \tilde{\gamma}_{2}}{2D} \sin \bigg(\frac{2 A}{\Omega} \cos \Omega t-2I_{0}t-2\varphi_{0}(0)\bigg) \nonumber \\ &-& \bigg (\frac{r \tilde{\gamma}_{2}}{2D}\bigg)^{2} \sin \bigg(\frac{3A}{\Omega} \cos \Omega t-3I_{0}t-3\varphi_{0}(0)\bigg), 
\label{eq:Is_phi_0_2}
\end{eqnarray}
and using $e^{i z \cos \theta} = \sum_{n=-\infty}^{\infty} i^{n} J_{n}(z) e^{ in \theta}$, $J_{n}(z)$ is the Bessel function of first kind, and n is integer, we come to
\begin{eqnarray}
I^{(0)}_{s} &=& Im\bigg\{\sum_{n=-\infty}^{\infty} i^{n}\bigg[ J_{n} \bigg(\frac{A}{\Omega}\bigg) \bigg[\bigg (\frac{r \tilde{\gamma}_{2}}{2D}\bigg)^{2}-1\bigg]e^{i((n \Omega -I_{0})t-\varphi_{0}(0))} + J_{n} \bigg(\frac{2A}{\Omega}\bigg)\frac{r \tilde{\gamma}_{2}}{2D} e^{i((n \Omega -2I_{0})t-2\varphi_{0}(0))} \nonumber \\ &-& J_{n} \bigg(\frac{3A}{\Omega}\bigg) \bigg (\frac{r \tilde{\gamma}_{2}}{2D}\bigg)^{2} e^{i((n \Omega -3I_{0})t-3\varphi_{0}(0)} \bigg]\bigg\}.
\label{eq:Is_phi_0_4s}
\end{eqnarray}
We see that Shapiro step appears when the ac component of the supercurrent vanishes. This means we have three possible cases: $I_{0}=n\Omega$, $I_{0}=n\Omega/2$, and $I_{0}=n\Omega/3$.\\
Now, for the first order we have
\begin{equation}
\dot{\varphi}_{1}=I_{1} +I_{s}
\end{equation}
which has the solution $<\dot{\varphi}_{1}>=0$ and $I_{1}=0$, so  $\dot{\varphi}_{1}(t)=I_{s}$. Next, we get the imaginary part of (\ref{eq:Is_phi_0_4s}) by separating odd and even  parts using
\begin{eqnarray}
i^{n} e^{i \theta} = i^{n}\cos \theta + i^{n+1} \sin \theta, \nonumber \\
\sum_{-\infty}^{\infty} i^{n} e^{i \theta}  = \sum_{-\infty}^{\infty} i^{2n} e^{i \theta} +\sum_{-\infty}^{\infty} i^{2n+1} e^{i \theta},
\label{eq:im_re}
\end{eqnarray}
with $i^{2n}=(-1)^{n}$, $Im\{i^{2n} e^{i\theta}\}=(-1)^{n}\sin \theta$, $Re\{i^{2n} e^{i\theta}\}=(-1)^{n}\cos \theta$,  $Im\{i^{2n+1} e^{i\theta}\}=(-1)^{n}\cos \theta$, and $Re\{i^{2n+1} e^{i\theta}\}=(-1)^{n+1}\sin \theta$. The imaginary part reads 
{\small
	\begin{eqnarray}
	\dot{\varphi}_{1}(t)&=& \sum_{n=-\infty}^{\infty} (-1)^{n}\bigg[ J_{2n} \bigg(\frac{A}{\Omega}\bigg) \bigg[1-\bigg(\frac{r \tilde{\gamma}_{2}}{2D}\bigg)^{2}\bigg]\sin(\Omega_{1}t-\varphi_{0}(0)) - J_{2n} \bigg(\frac{2A}{\Omega}\bigg)\frac{r \tilde{\gamma}_{2}}{2D} \sin(\Omega_{2}t-2\varphi_{0}(0)) \nonumber \\ &+& J_{2n} \bigg(\frac{3A}{\Omega}\bigg) \bigg (\frac{r \tilde{\gamma}_{2}}{2D}\bigg)^{2} \sin(\Omega_{3}t-3\varphi_{0}(0))+J_{2n+1} \bigg(\frac{A}{\Omega}\bigg) \bigg[1-\bigg(\frac{r \tilde{\gamma}_{2}}{2D}\bigg)^{2}\bigg]\cos(\tilde{\Omega}_{1}t-\varphi_{0}(0)) \nonumber \\&-& J_{2n+1} \bigg(\frac{2A}{\Omega}\bigg)\frac{r \tilde{\gamma}_{2}}{2D} \cos(\tilde{\Omega}_{2}t-2\varphi_{0}(0)) + J_{2n+1} \bigg(\frac{3A}{\Omega}\bigg) \bigg (\frac{r \tilde{\gamma}_{2}}{2D}\bigg)^{2} \cos(\tilde{\Omega}_{3}t-3\varphi_{0}(0)) \bigg],
	\label{eq:Is_phi_0_5}
	\end{eqnarray}}
where $\Omega_{1}=2n \Omega -I_{0}$, $\Omega_{2}=2n \Omega -2I_{0}$, $\Omega_{3}=2n \Omega -3I_{0}$, $\tilde{\Omega}_{1}=(2n+1) \Omega -I_{0}$, $\tilde{\Omega}_{2}=(2n+1) \Omega -2I_{0}$, and $\tilde{\Omega}_{3}=(2n+1) \Omega -3I_{0}$. By integrating Eq.(\ref{eq:Is_phi_0_5}) with respect to time, we get
{\small
	{
		\allowdisplaybreaks
		\begin{eqnarray}
		\varphi_{1}(t)&=& \sum_{n=-\infty}^{\infty} (-1)^{n+1}\bigg[  J_{2n} \bigg(\frac{A}{\Omega}\bigg) \bigg[1-\bigg (\frac{r \tilde{\gamma}_{2}}{2D}\bigg)^{2}\bigg]\frac{\cos(\Omega_{1}t-\varphi_{0}(0))}{\Omega_{1}} -J_{2n} \bigg(\frac{2A}{\Omega}\bigg)\frac{r \tilde{\gamma}_{2}}{2D} \frac{\cos(\Omega_{2}t-2\varphi_{0}(0))}{\Omega_{2}} \nonumber \\ &+& J_{2n} \bigg(\frac{3A}{\Omega}\bigg) \bigg (\frac{r \tilde{\gamma}_{2}}{2D}\bigg)^{2} \frac{\cos(\Omega_{3}t-3\varphi_{0}(0))}{\Omega_{3}} \bigg]+(-1)^{n}\bigg[J_{2n+1} \bigg(\frac{A}{\Omega}\bigg) \bigg[1-\bigg (\frac{r \tilde{\gamma}_{2}}{2D}\bigg)^{2}\bigg]\frac{\sin(\tilde{\Omega}_{1}t-\varphi_{0}(0))}{\tilde{\Omega}_{1}} \nonumber \\&-& J_{2n+1} \bigg(\frac{2A}{\Omega}\bigg)\frac{r \tilde{\gamma}_{2}}{2D} \frac{\sin(\tilde{\Omega}_{2}t-2\varphi_{0}(0))}{\tilde{\Omega}_{2}} + J_{2n+1} \bigg(\frac{3A}{\Omega}\bigg) \bigg (\frac{r \tilde{\gamma}_{2}}{2D}\bigg)^{2} \frac{\sin(\tilde{\Omega}_{3}t-3\varphi_{0}(0))}{\tilde{\Omega}_{3}} \bigg].
		\label{eq:Is_phi_1_6}
		\end{eqnarray}}}
For the $1^{st}$ order,  the supercurrent is given by ($	I^{(1)}_{s}=-	f_{2}(t)$)
\begin{eqnarray}
I^{(1)}_{s}&=&\bigg[1-\bigg (\frac{r \tilde{\gamma}_{2}}{2D}\bigg)^{2}\bigg] \varphi_{1}(t)\cos\varphi_{0}(t) -\frac{r \tilde{\gamma}_{2}}{2D} 2\varphi_{1}(t)\cos 2\varphi_{0}(t) +\bigg (\frac{r \tilde{\gamma}_{2}}{2D}\bigg)^{2} 3\varphi_{1}(t)\cos 3\varphi_{0}(t). \
\label{eq:Is_phi_1_7}
\end{eqnarray}
Now, we need to insert Eq.(\ref{eq:Is_phi_1_6}) into Eq.(\ref{eq:Is_phi_1_7}) using
\begin{eqnarray}
\cos (q \varphi_{0}(t)) &=& \sum_{m=-\infty}^{\infty} J_{2m} \bigg(\frac{A}{\Omega}\bigg)\cos(q(\Omega_{m}-\varphi(0)_{0})) + J_{2m+1} \bigg(\frac{A}{\Omega}\bigg)\sin(q(\tilde{\Omega}_{m}-\varphi(0)_{0})),
\end{eqnarray}
where $q=1,2,3$, $\Omega_{m}=2m\Omega-I_{0}$,$\tilde{\Omega}_{m}=(2m+1)\Omega-I_{0}$ , and m is integer. The final expression for $I_{s}$ is given by
\begin{eqnarray}
I^{(1)}_{s}&=&\bigg[1-\bigg (\frac{r \tilde{\gamma}_{2}}{2D}\bigg)^{2}\bigg]  \varGamma^{(1)} -\frac{r \tilde{\gamma}_{2}} {D} \varGamma^{(2)} + 3\bigg (\frac{r \tilde{\gamma}_{2}}{2D}\bigg)^{2} \varGamma^{(3)},
\label{eq:Is_phi_1_8}
\end{eqnarray}
with  \\ \\ \\
{\small
	{
		\allowdisplaybreaks
		\begin{eqnarray}
		\varGamma^{(k)}&=&\sum_{n,m}(-1)^{n+m+1}\bigg\{\bigg[\bigg (\frac{r \tilde{\gamma}_{2}}{2D}\bigg)^{2}-1\bigg]  \frac{B^{1}_{2n,2m}}{2 \Omega_{1}} \bigg[\cos((\Omega_{1}+k\Omega_{m})t-(k+1)\varphi_{0}(0))\nonumber \\ &+&\cos((\Omega_{1}-k\Omega_{m})t-(k-1)\varphi_{0}(0))\bigg]\nonumber -\frac{r \tilde{\gamma}_{2}} {2D} \frac{B^{2}_{2n,2m}}{2 \Omega_{2}} \bigg[\cos((\Omega_{2}+k\Omega_{m})t-(k+2)\varphi_{0}(0))\nonumber \\ &+&\cos((\Omega_{2}-k\Omega_{m})t-k\varphi_{0}(0))\bigg] + \bigg (\frac{r \tilde{\gamma}_{2}}{2D}\bigg)^{2} \frac{B^{3}_{2n,2m}}{2 \Omega_{3}} \bigg[\cos((\Omega_{3}+k\Omega_{m})t-(k+3)\varphi_{0}(0))\nonumber\\&+&\cos((\Omega_{3}-k\Omega_{m})t-(k+1)\varphi_{0}(0))\bigg]\bigg\}\nonumber\\&+&
		(-1)^{n+m}\bigg\{\bigg[\bigg (\frac{r \tilde{\gamma}_{2}}{2D}\bigg)^{2}-1\bigg]  \frac{B^{1}_{2n+1,2m}}{2 \tilde{\Omega}_{1}} \bigg[\sin((\tilde{\Omega}_{1}+k\Omega_{m})t-(k+1)\varphi_{0}(0))\nonumber \\ &+&\sin((\tilde{\Omega}_{1}-k\Omega_{m})t-(k-1)\varphi_{0}(0))\bigg]-\frac{r \tilde{\gamma}_{2}} {2D} \frac{B^{2}_{2n+1,2m}}{2 \tilde{\Omega}_{2}} \bigg[\sin((\tilde{\Omega}_{2}+k\Omega_{m})t-(k+2)\varphi_{0}(0))\nonumber\\&+&\sin((\tilde{\Omega}_{2}-k\Omega_{m})t-k\varphi_{0}(0))\bigg]+ \bigg (\frac{r \tilde{\gamma}_{2}}{2D}\bigg)^{2} \frac{B^{3}_{2n+1,2m}}{2 \tilde{\Omega}_{3}} \bigg[\sin((\tilde{\Omega}_{3}+k\Omega_{m})t-(k+3)\varphi_{0}(0))\nonumber\\&+&\sin((\tilde{\Omega}_{3}-k\Omega_{m})t-(k+1)\varphi_{0}(0))\bigg]\bigg\}\nonumber\\&+&
		(-1)^{n+m+2}\bigg\{\bigg[\bigg (\frac{r \tilde{\gamma}_{2}}{2D}\bigg)^{2}-1\bigg]  \frac{B^{1}_{2n,2m+1}}{2 \Omega_{1}} \bigg[\sin((\Omega_{1}+k\tilde{\Omega}_{m})t-(k+1)\varphi_{0}(0))\nonumber\\&+&\sin((\Omega_{1}-k\tilde{\Omega}_{m})t-(k-1)\varphi_{0}(0))\bigg]-\frac{r \tilde{\gamma}_{2}} {2D} \frac{B^{2}_{2n,2m+1}}{2 \Omega_{2}} \bigg[\sin((\Omega_{2}+k\tilde{\Omega}_{m})t-(k+2)\varphi_{0}(0))\nonumber\\&+&\sin((\Omega_{2}-k\tilde{\Omega}_{m})t-k\varphi_{0}(0))\bigg] + \bigg (\frac{r \tilde{\gamma}_{2}}{2D}\bigg)^{2} \frac{B^{3}_{2n,2m+1}}{2 \Omega_{3}} \bigg[\sin((\Omega_{3}+k\tilde{\Omega}_{m})t-(k+3)\varphi_{0}(0))\nonumber\\&+&\sin((\Omega_{3}-k\tilde{\Omega}_{m})t-(k+1)\varphi_{0}(0))\bigg]\bigg\}\nonumber\\&+&
		(-1)^{n+m}\bigg\{\bigg[\bigg (\frac{r \tilde{\gamma}_{2}}{2D}\bigg)^{2}-1\bigg]  \frac{B^{1}_{2n+1,2m+1}}{2 \tilde{\Omega}_{1}} \bigg[\cos((\tilde{\Omega}_{1}+k\tilde{\Omega}_{m})t-(k+1)\varphi_{0}(0))\nonumber\\&+&\cos((\tilde{\Omega}_{1}-k\tilde{\Omega}_{m})t-(k-1)\varphi_{0}(0))\bigg]-\frac{r \tilde{\gamma}_{2}} {2D} \frac{B^{2}_{2n+1,2m}}{2 \tilde{\Omega}_{2}} \bigg[\cos((\tilde{\Omega}_{2}+k\tilde{\Omega}_{m})t-(k+2)\varphi_{0}(0))\nonumber\\&+&\cos((\tilde{\Omega}_{2}-k\tilde{\Omega}_{m})t-k\varphi_{0}(0))\bigg] + \bigg (\frac{r \tilde{\gamma}_{2}}{2D}\bigg)^{2} \frac{B^{3}_{2n+1,2m}}{2 \tilde{\Omega}_{3}} \bigg[\cos((\tilde{\Omega}_{3}+k\tilde{\Omega}_{m})t-(k+3)\varphi_{0}(0))\nonumber\\&+&\cos((\tilde{\Omega}_{3}-k\tilde{\Omega}_{m})t-(k+1)\varphi_{0}(0))\bigg]\bigg\},
		\label{eq:Is_phi_1_9}
		\end{eqnarray}
}}
where $k$ takes the value of 1,2,3, $B^{1}_{x,y}$=$J_{x}(A/\Omega)J_{y}(A/\Omega)$,$B^{2}_{x,y}$=$J_{x}(2A/\Omega)J_{y}(A/\Omega)$ and $B^{3}_{x,y}=J_{x}(3A/\Omega)J_{y}(A/\Omega)$. The ac component of the supercurrent vanishes at $\Omega_{1}=\pm k\Omega_{m}$, $\Omega_{2}=\pm k\Omega_{m}$, $\Omega_{3}=\pm k\Omega_{m}$,$\tilde{\Omega}_{1}=\pm k\Omega_{m}$, $\tilde{\Omega}_{2}=\pm k\Omega_{m}$, $\tilde{\Omega}_{3}=\pm k\Omega_{m}$, $\Omega_{1}=\pm k\tilde{\Omega}_{m}$, $\Omega_{2}=\pm k\tilde{\Omega}_{m}$, $\Omega_{3}=\pm k\tilde{\Omega}_{m}$,$\tilde{\Omega}_{1}=\pm k\tilde{\Omega}_{m}$, $\tilde{\Omega}_{2}=\pm k\tilde{\Omega}_{m}$, $\tilde{\Omega}_{3}=\pm k\tilde{\Omega}_{m}$. 
\begin{figure}[h]
	\centering
	\includegraphics[width=0.48\linewidth, angle=0]{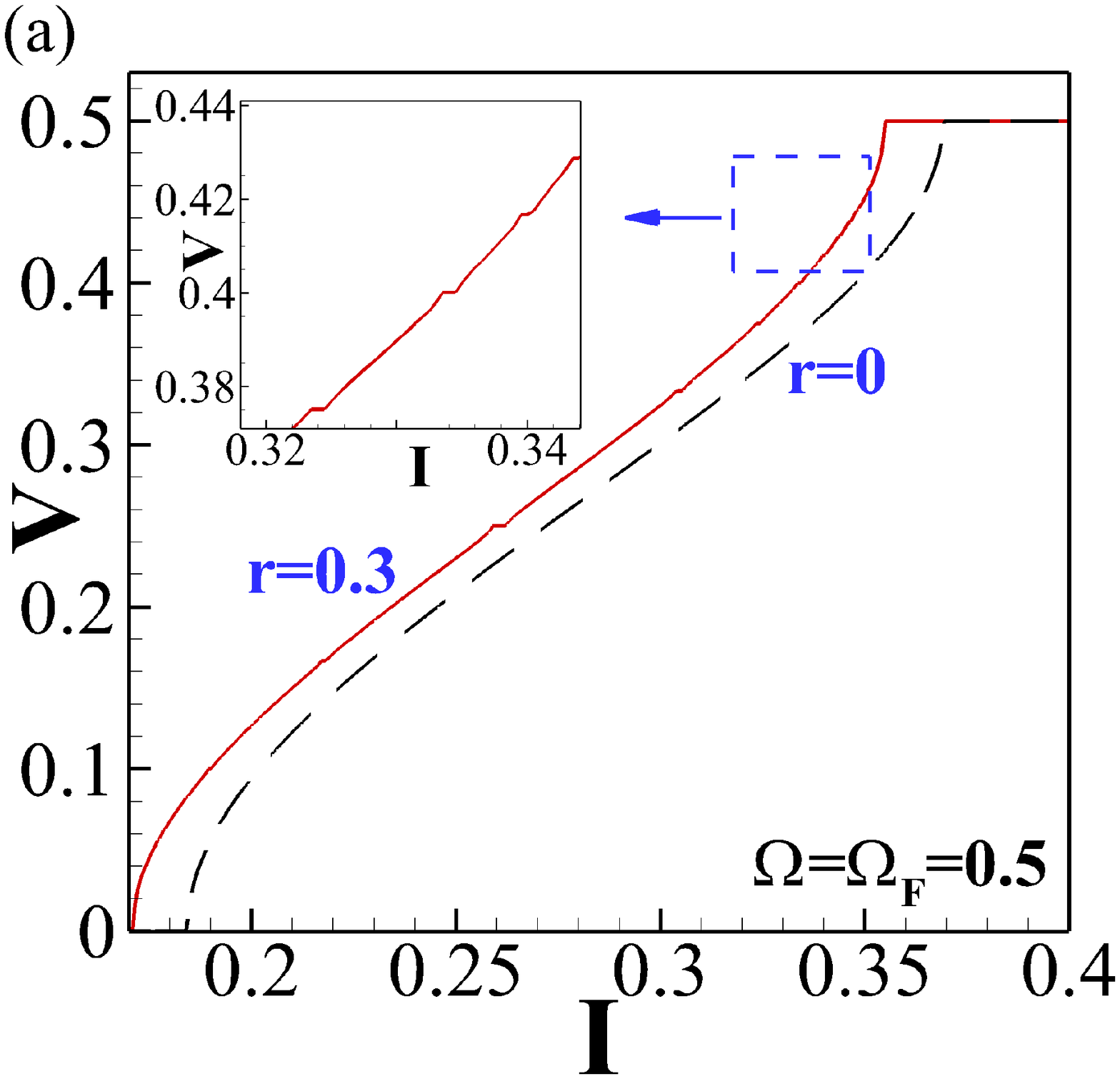}
	\includegraphics[width=0.48\linewidth, angle=0]{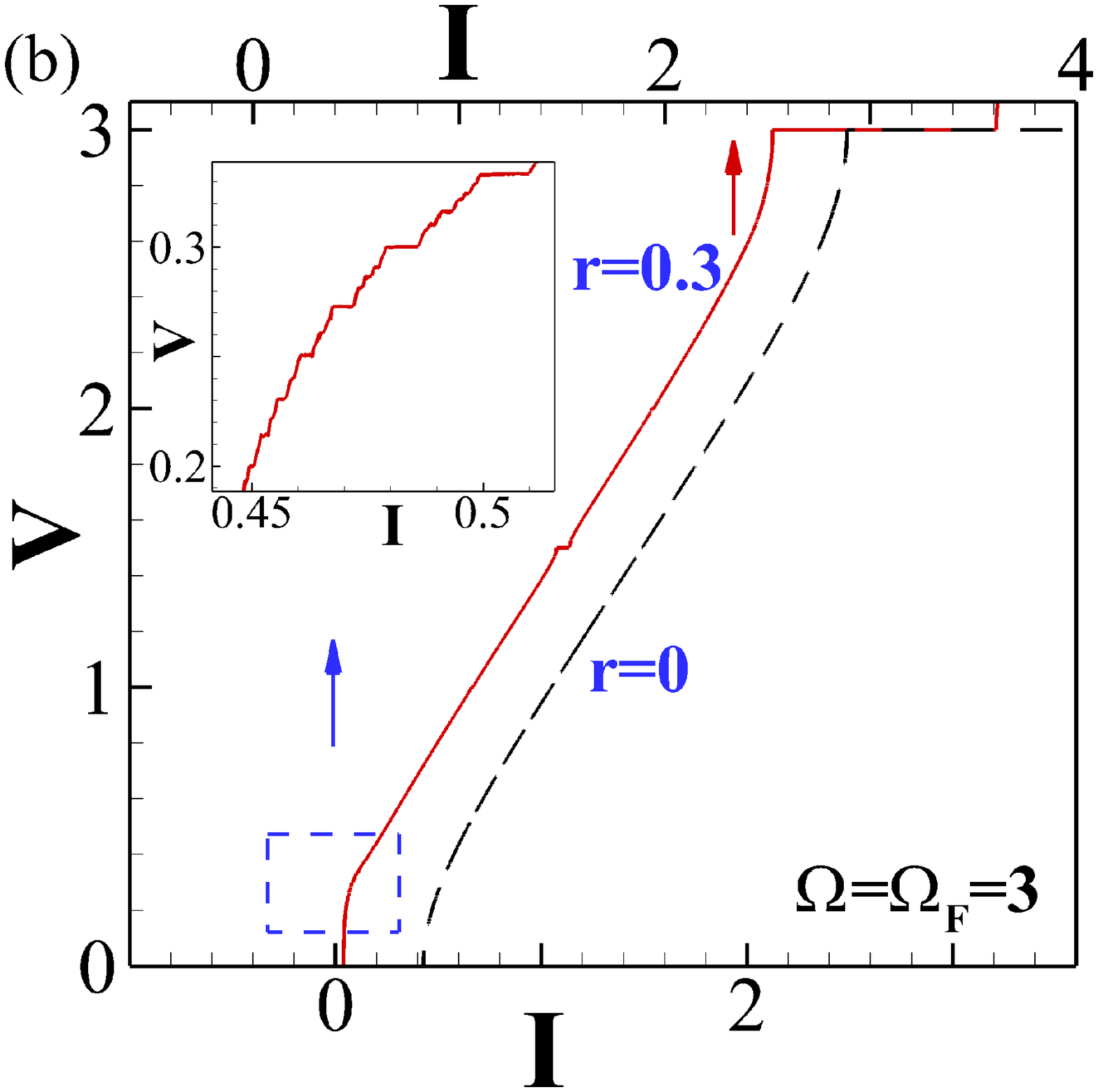}
	\caption{ parts of IV-characteristic for overdamped $\varphi_{0}$-junction without ($r=0$) and with spin orbit coupling at (a) $\Omega=\Omega_{F}=0.5$, (b) $\Omega=\Omega_{F}=3$. For all figures we take $\alpha=0.1$, $G=0.6$, and $A=5$.  }
	\label{1s}
\end{figure}

Finally,  we confirm the appearance of subharmonic steps which are predicted from the perturbative analysis. This perturbative analysis can be applied when at least one of these condition is satisfied $A>>1$, or $\Omega>>1$, or $\beta_{c}\Omega^{2}>>1$ \cite{Kornev}. 

In figures \ref{1s}(a) and (b) we show an analogous plot for the IV-characteristics of JJ without and with spin-orbit coupling ($r=0.3$) at $A=5$. As it is predicted by the perturbative analysis, a set of subharmonic steps appear when $r \neq 0 $ (see figures \ref{1s}(a) and (b)).

\end{widetext}	
	
\end{document}